\documentclass[12pt]{article}
\usepackage{amssymb}
\usepackage{graphicx}
\begin{document}
\newcommand{\beq}{\begin{equation}}
\newcommand{\eeq}{\end{equation}}
\newcommand{\beqa}{\begin{eqnarray}}
\newcommand{\eeqa}{\end{eqnarray}}
\newcommand{\beqar}{\begin{eqnarray*}}
\newcommand{\eeqar}{\end{eqnarray*}}
\newcommand{\al}{\alpha}
\newcommand{\be}{\beta}
\newcommand{\del}{\delta}
\newcommand{\D}{\Delta}
\newcommand{\eps}{\epsilon}
\newcommand{\ga}{\gamma}
\newcommand{\Ga}{\Gamma}
\newcommand{\ka}{\kappa}
\newcommand{\nn}{\nonumber}
\newcommand{\inn}{\!\cdot\!}
\newcommand{\h}{\eta}
\newcommand{\ii}{\iota}
\newcommand{\kk}{\varphi}
\newcommand\F{{}_3F_2}
\newcommand{\la}{\lambda}
\newcommand{\La}{\Lambda}
\newcommand{\na}{\prt}
\newcommand{\Om}{\Omega}
\newcommand{\om}{\omega}
\newcommand{\p}{\Phi}
\newcommand{\sig}{\sigma}
\renewcommand{\t}{\theta}
\newcommand{\z}{\zeta}
\newcommand{\ssc}{\scriptscriptstyle}
\newcommand{\eg}{{\it e.g.,}\ }
\newcommand{\ie}{{\it i.e.,}\ }
\newcommand{\labell}[1]{\label{#1}} %{\label{#1}} %
\newcommand{\reef}[1]{(\ref{#1})}
\newcommand\prt{\partial}
\newcommand\veps{\varepsilon}
\newcommand{\pol}{\varepsilon}
\newcommand\vp{\varphi}
\newcommand\ls{\ell_s}
\newcommand\cF{{\cal F}}
\newcommand\cA{{\cal A}}
\newcommand\cS{{\cal S}}
\newcommand\cT{{\cal T}}
\newcommand\cV{{\cal V}}
\newcommand\cL{{\cal L}}
\newcommand\cM{{\cal M}}
\newcommand\cN{{\cal N}}
\newcommand\cG{{\cal G}}
\newcommand\cK{{\cal K}}
\newcommand\cH{{\cal H}}
\newcommand\cI{{\cal I}}
\newcommand\cJ{{\cal J}}
\newcommand\cl{{\iota}}
\newcommand\cP{{\cal P}}
\newcommand\cQ{{\cal Q}}
\newcommand\cg{{\tilde {{\cal G}}}}
\newcommand\cR{{\cal R}}
\newcommand\cB{{\cal B}}
\newcommand\cO{{\cal O}}
\newcommand\tcO{{\tilde {{\cal O}}}}
\newcommand\bz{\bar{z}}
\newcommand\bb{\bar{b}}
\newcommand\ba{\bar{a}}
\newcommand\bg{\bar{g}}
\newcommand\bc{\bar{c}}
\newcommand\bw{\bar{w}}
\newcommand\bX{\bar{X}}
\newcommand\bK{\bar{K}}
\newcommand\bA{\bar{A}}
\newcommand\bZ{\bar{Z}}
\newcommand\bxi{\bar{\xi}}
\newcommand\bphi{\bar{\phi}}
\newcommand\bpsi{\bar{\psi}}
\newcommand\bprt{\bar{\prt}}
\newcommand\bet{\bar{\eta}}
\newcommand\btau{\bar{\tau}}
\newcommand\hF{\hat{F}}
\newcommand\hA{\hat{A}}
\newcommand\hT{\hat{T}}
\newcommand\htau{\hat{\tau}}
\newcommand\hD{\hat{D}}
\newcommand\hf{\hat{f}}
\newcommand\hK{\hat{K}}
\newcommand\hg{\hat{g}}
\newcommand\hp{\hat{\Phi}}
\newcommand\hi{\hat{i}}
\newcommand\ha{\hat{a}}
\newcommand\hb{\hat{b}}
\newcommand\hQ{\hat{Q}}
\newcommand\hP{\hat{\Phi}}
\newcommand\hS{\hat{S}}
\newcommand\hX{\hat{X}}
\newcommand\tL{\tilde{\cal L}}
\newcommand\hL{\hat{\cal L}}
\newcommand\tG{{\tilde G}}
\newcommand\tg{{\tilde g}}
\newcommand\tphi{{\widetilde \Phi}}
\newcommand\tPhi{{\widetilde \Phi}}
\newcommand\te{{\tilde e}}
\newcommand\tk{{\tilde k}}
\newcommand\tf{{\tilde f}}
\newcommand\ta{{\tilde a}}
\newcommand\tb{{\tilde b}}
\newcommand\tc{{\tilde c}}
\newcommand\td{{\tilde d}}
\newcommand\tm{{\tilde m}}
\newcommand\tmu{{\tilde \mu}}
\newcommand\tnu{{\tilde \nu}}
\newcommand\talpha{{\tilde \alpha}}
\newcommand\tbeta{{\tilde \beta}}
\newcommand\trho{{\tilde \rho}}
 \newcommand\tR{{\tilde R}}
\newcommand\teta{{\tilde \eta}}
\newcommand\tF{{\widetilde F}}
\newcommand\tK{{\tilde K}}
\newcommand\tE{{\widetilde E}}
\newcommand\tpsi{{\tilde \psi}}
\newcommand\tX{{\widetilde X}}
\newcommand\tD{{\widetilde D}}
\newcommand\tO{{\widetilde O}}
\newcommand\tS{{\tilde S}}
\newcommand\tB{{\tilde B}}
\newcommand\tA{{\widetilde A}}
\newcommand\tT{{\widetilde T}}
\newcommand\tC{{\widetilde C}}
\newcommand\tV{{\widetilde V}}
\newcommand\thF{{\widetilde {\hat {F}}}}
\newcommand\Tr{{\rm Tr}}
\newcommand\tr{{\rm tr}}
\newcommand\STr{{\rm STr}}
\newcommand\hR{\hat{R}}
\newcommand\M[2]{M^{#1}{}_{#2}}
\newcommand\MZ{\mathbb{Z}}
\newcommand\MR{\mathbb{R}}
\newcommand\bS{\textbf{ S}}
\newcommand\bI{\textbf{ I}}
\newcommand\bJ{\textbf{ J}}

%\begin{document}
\begin{titlepage}
\begin{center}

\vskip 2 cm
{\LARGE \bf  
D-Brane Effective Lagrangian in \\  \vskip 0.25 cm Spacetimes with Boundaries
 }\\
\vskip 1.25 cm
  Ali Baradaran-Hosseini\footnote{ali.baradaran.hosseini.2@gmail.com} and  Mohammad R. Garousi\footnote{garousi@um.ac.ir}

\vskip 1 cm
{{\it Department of Physics, Faculty of Science, Ferdowsi University of Mashhad\\}{\it P.O. Box 1436, Mashhad, Iran}\\}
\vskip .1 cm
 \end{center}

\begin{abstract}

In this study, we explore the transformation of $D_p$-branes to $D_{p-1}$-branes under T-duality when the $D$-brane is embedded in a spacetime with a boundary. Our goal is to derive the higher-derivative corrections to the Dirac-Born-Infeld (DBI) Lagrangian for both the bulk and boundary terms.

For the bulk terms, we calculate the $\alpha'$ corrections for the massless open string fields, up to the 8th order in the dimensionless Maxwell field strength. We demonstrate that the bulk Lagrangian can satisfy the T-duality constraint without residual total derivative terms in the base space. This determines the most general independent couplings of the massless open string fields in the bulk Lagrangian, encompassing 145 coupling constants, up to three parameters. Two of these parameters are physical and are determined by disk-level S-matrix elements, while the third is unphysical and can be eliminated by field redefinitions and integration by parts. The final result for the bulk Lagrangian consists of 49 couplings.

For the boundary terms, applying T-duality symmetry to the massless open string field allows us to extend the DBI Lagrangian to incorporate the extrinsic curvature of the boundary.
\end{abstract}

%Keywords: T-duality, D-brane effective action
\end{titlepage}

\section{Introduction}

The effective spacetime Lagrangian of string theory on a background $\mathcal{M}^{(D)}$ with a codimension-1 boundary $\partial\mathcal{M}^{(D)}$ exhibits an infinite expansion in powers of the string tension parameter $\alpha'$ for both the bulk and boundary Lagrangian terms. Similarly, the effective world-volume Lagrangian of the non-perturbative $D_p$-brane objects in this background also displays an infinite $\alpha'$ expansion for their bulk and boundary world-volume Lagrangian terms. These effective Lagrangians can be derived by exploring the various spacetime symmetries inherent to string theory or by using the world-sheet non-linear sigma model. The resulting expressions must also be consistent with the principle of stationary action, subject to the appropriate boundary conditions.

One of the most important symmetries that appears in all string theories, after compactifying them on the torus $T^d$, is T-duality \cite{Giveon:1994fu,Alvarez:1994dn}. It has been argued \cite{Sen:1991zi,Hohm:2014sxa} that the dimensional reduction of the classical effective action of string theory on tori are invariant under $O(d,d,\mathbb{R})$ transformations at all orders of $\alpha'$.
For circular reduction, a non-geometric subgroup of $O(1,1,\mathbb{R})$ which is a $\mathbb{Z}_2$-group may be used to construct the covariant effective action or effective Lagrangian at any order of $\alpha'$ in terms of a few unfixed parameters \cite{Garousi:2019wgz,Garousi:2019mca,Garousi:2020gio}. For the torus reduction, the $O(d,d,\mathbb{R})$ symmetry excludes some covariant couplings in the base space. This constraint may also be used in determining the covariant couplings in the parent theory \cite{Wulff:2024ips,Wulff:2024mgu}.
Furthermore, a non-geometric aspect of $O(D,D,\mathbb{R})$ symmetry, known as $\beta$-transformation, has been employed in \cite{Baron:2022but,Garousi:2022qmk,Garousi:2023diq,Baron:2023qkx} to determine the coupling constants without relying on dimensional reduction.

The circular reduction of non-perturbative $D_p$-brane objects transforms covariantly under the $\mathbb{Z}_2$-group. When a $D_p$-brane is along a circle, the world-volume reduction of its Lagrangian should transform under T-duality to the transverse reduction of a $D_{p-1}$-brane Lagrangian \cite{Bergshoeff:1996cy}. This is provided that the circular reduction does not break the diffeomorphism symmerty of the parent couplings. In other words, the circular reduction must have a $U(1) \times U(1)$ gauge symmetry, where the first $U(1)$ corresponds to the momentum vector and the second $U(1)$ corresponds to the winding momentum. This symmetry always exists for the circular reduction of the covariant spacetime actions, whereas it may not exist for the circular reduction of covariant $D_p$-brane actions in the static gauge for a general base space background \cite{Hosseini:2022vrr,Garousi:2024rzh}.

Assuming that the coupling constants in the effective Lagrangians of string theory at the critical dimension are background independent \cite{Hohm:2016lge,Garousi:2022ovo}, we can consider a particular background with a circle in both the bulk and the boundary, i.e., $\mathcal{M}^{(D)}=\mathcal{M}^{(D-1)}\times S^{(1)}$ and $\partial\mathcal{M}^{(D)}=\mathcal{M}^{(D-2)}\times S^{(1)}$. The coupling constants for this background, which can be found by T-duality, should be the same as the coupling constants for any other background. The background independence also dictates that both the bulk and boundary Lagrangians should separately satisfy the T-duality constraints. The bulk Lagrangian should be the same for a spacetime with or without a boundary. Both should be invariant under T-duality. The T-duality of the Lagrangian in spacetime with a boundary then dictates that the boundary Lagrangian should also be invariant under T-duality. This T-duality symmetry can be imposed as a constraint on the most general independent covariant and gauge-invariant Lagrangian with arbitrary coupling constants. Requiring the Lagrangian to be invariant under the $\mathbb{Z}_2$ transformation can fix or determine the arbitrary coupling constants.

When the spacetime has no boundary, the T-duality transformations can be applied to the effective actions, and the total derivative terms can be safely ignored. In this case, the T-duality constraint has been used to fix the Neveu-Schwarz-Neveu-Schwarz (NS-NS) couplings up to order $\alpha'^3$ \cite{Garousi:2019wgz,Garousi:2019mca,Garousi:2020gio} and to fix the  NS-NS and Yang-Mills (YM) couplings in the heterotic theory up to order $\alpha'^2$ \cite{Garousi:2024avb,Garousi:2024vbz,Garousi:2023kxw,Garousi:2024imy}.
Building on the above arguments, for the case where the spacetime has a boundary, one expects that by adding certain total derivative terms to the effective actions found in \cite{Garousi:2019wgz,Garousi:2019mca,Garousi:2020gio,Garousi:2024avb,Garousi:2024vbz,Garousi:2023kxw,Garousi:2024imy}, one should be able to extend them to bulk Lagrangians that are invariant under T-duality without ignoring any total derivative terms in the base space. Such Lagrangians at order $\alpha'$ have been found in \cite{Garousi:2023qwj}.
Then, one expects the boundary Lagrangians at any order of $\alpha'$ to be also separately invariant under the T-duality transformation. This constraint may fix the boundary couplings as well. The final bulk and boundary Lagrangian should be consistent with the principle of stationary action, subject to the appropriate boundary conditions. 

At the leading order of $\alpha'$, it has been shown in \cite{Becker:2010ij} that the following Lagrangian is invariant under T-duality:
\beqa
\bold{L}^{(0)}&=& -\frac{2}{\kappa^2} e^{-2\Phi}\sqrt{-G}\,  \left( R -4\nabla_{\mu}\Phi \nabla^{\mu}\Phi-\frac{1}{12} H^2+4\nabla_\mu\nabla^\mu\Phi\right)\,,\labell{S0b}
\eeqa
which is the leading order effective action of string theory, up to a total derivative term. At the leading order of $\alpha'$, it has been shown in \cite{Garousi:2019xlf} that the following boundary Lagrangian is also invariant under the T-duality transformation:
\beqa
\partial\bold{L}^{(0)}&=&  -\frac{a}{\kappa^2}\, e^{-2\Phi}\sqrt{|g|}\,  \left(- K+2n^{\mu}\nabla_{\mu}\Phi \right)\,,\labell{bbaction}
\eeqa
where $a$ is an unfixed parameter, $n_\mu$ is the unit vector, $n_\mu n^\mu=1$, normal to the boundary, and $K$ is the trace of the extrinsic curvature of the boundary
\beqa
 K_{\mu\nu}&=&\nabla_\mu n_\nu-n_\mu n^\rho\nabla_\rho n_\nu\,.\labell{ext}
\eeqa
The boundary couplings \reef{bbaction} are consistent with the principle of stationary action under Dirichlet boundary conditions when $a=-4$ \cite{Kazakov:2001pj}. The same number $a=-4$ can also be found by imposing the condition that the combination of bulk and boundary actions is invariant under the T-duality \cite{Garousi:2019xlf}. A similar idea has been used in \cite{Ahmadain:2024uom} to find the boundary term by the world-sheet non-linear sigma model and found one-half of the Gibbons-Hawking-York term \cite{York:1972sj,Gibbons:1976ue}.  
At higher orders of $\alpha'$, one should impose appropriate boundary conditions on the massless fields and their derivatives, as well as appropriate field redefinitions, in order to satisfy the principle of stationary action \cite{Garousi:2021cfc,Garousi:2021yyd}. However, we are not interested in exploring these  higher-order considerations in the present work.

In this paper, we aim to study the bulk and boundary world-volume Lagrangians for $D_p$-branes. The reason that both the bulk and boundary world-volume Lagrangians should separately satisfy the T-duality constraint is that we may consider the particular case where the $D_p$-brane is entirely along the boundary, i.e., $\partial\mathcal{M}^{(D)}=\mathcal{M}^{(p+1)}\times \mathcal{M}^{(D-p-2)}$ and the $D_p$-brane is along the subspace $\mathcal{M}^{(p+1)}$. In this case, there is only a boundary world-volume Lagrangian, which must still satisfy the T-duality constraint. Hence, for the general case where the $D_p$-brane is extended in the bulk and ends on the boundary, requiring both the bulk and boundary Lagrangians to separately satisfy T-duality is necessary.
In \cite{Akou:2020mxx,Hosseini:2022vrr}, the T-duality constraint has been applied to the combination of bulk and boundary world-volume actions to determine specific world-volume boundary couplings.

The bulk Lagrangian at leading order in $\alpha'$  is given by the DBI Lagrangian \cite{Leigh:1989jq,Bachas:1995kx}
\beqa
\bold{L}_p^{(0)}&=&-T_p\sqrt{-\det(\tG_{ab}+F_{ab})}\, ,\labell{DBI}
\eeqa
where $T_p$ is the tension of the $D_p$-brane, $F_{ab}$ is the field strength of the Maxwell gauge field $A_a(\sigma)$, and $\tG_{ab}$ is the pull-back of the bulk metric onto the world-volume, i.e.,
\beqa
\tG_{ab}=\prt_a X^\mu\prt_b X^\nu G_{\mu\nu}\,,\labell{pull}
\eeqa
where $X^\mu(\sigma)$ is the spacetime coordinate that specifies the $D_p$-brane in the spacetime, and $G_{\mu\nu}$ is the spacetime metric that we assume to be diagonal and constant in this Lagrangian.
We also assume the $B$-field and dilaton are zero in this Lagrangian, hence, the circular reduction has no momentum and winding gauge fields, i.e., its circular reduction has a $U(1) \times U(1)$ symmetry. Since the gauge field $A_a$ and the world-volume field $X^\mu$ are related to each other by the T-duality transformation, we have normalized the gauge field $A_a$ to have the same dimension as the world-volume field $X^\mu$. With this normalization, the action above is at leading order in $\alpha'$.
The world-volume reduction of the above $D_p$-brane Lagrangian transforms to the transverse reduction of the $D_{p-1}$-brane Lagrangian under T-duality transformations (see e.g. \cite{Myers:1999ps})\footnote{ 
To ensure full symmetry, including the brane tension, the corresponding action should be invariant under T-duality. However, in this paper, we will not employ any integration by parts. }.  The $\alpha'$ corrections to the above DBI action  have been studied in \cite{Abouelsaood:1986gd,Tseytlin:1987ww,Andreev:1988cb, Wyllard:2000qe, Andreev:2001xx, Wyllard:2001ye,Karimi:2018vaf,Garousi:2022rcv}.

There is an infinite number of $F$ terms in the $\alpha'$ corrections to the above Lagrangian. The corrections up to the 8th order in $F$ for the bulk effective action, ignoring total derivative terms, have been found in \cite{Karimi:2018vaf} by T-duality.
In deriving these results, one first removes the redundant terms due to $X^\mu$- and $A_a$-field redefinitions, as well as integration by parts, from the most general gauge-invariant couplings to find an independent basis with arbitrary coupling constants. Then, one uses T-duality to fix the coupling constants up to two unfixed parameters. The two parameters are also fixed by the disk-level four-point function. It turns out that by adding some total derivative terms to the effective world-volume action found in \cite{Karimi:2018vaf}, one cannot find the effective world-volume Lagrangian that would be invariant under the T-duality. This is unlike the spacetime effective Lagrangians that can be found by adding some total derivative terms to the spacetime effective action. However, the world-volume Lagrangian should also satisfy the T-duality constraint.

To find the world-volume Lagrangian  using T-duality, we observe that T-duality is satisfied only by the basis in which the terms related to each other by $X^\mu$-field redefinition are not removed.
In this paper, among other things, we will find the effective Lagrangian of massless open string fields up to 8th order in $F$ by first finding the basis in which the redundant terms from only the gauge field redefinitions are removed, and then imposing T-duality to determine their corresponding coupling constants. We find that the couplings are fixed up to three parameters. One of them can be removed by field redefinition and integration by parts, and the other two parameters are physical which can be fixed by disk-level four-point S-matrix element or by comparing with the effective action found in \cite{Karimi:2018vaf}.

To explore the boundary world-volume Lagrangian, we examine the case where the $D_p$-brane is entirely along the boundary. In this setup, there exists an additional vector field $n_a = \partial_a X^\mu n_\mu$ in addition to the massless fields in the DBI Lagrangian. Our objective is to incorporate this vector field and its first derivative in the DBI Lagrangian. We will demonstrate that the T-duality of the massless open  string fields determines the following boundary Lagrangian:
\beqa
\prt\bold{L}_p&=&-T_p\, e^{-\Phi}\sqrt{-\det(\tG_{ab}+\tB_{ab}+F_{ab}+2\sqrt{\alpha'}\prt_a X^\mu\prt_b X^\nu K_{\mu\nu})}\,,\labell{DBIK}
\eeqa
%In proving the T-duality of the above Lagrangian, we consider a particular background for the massless closed string fields, where the circular reduction respects the $U(1) \times U(1)$ symmetry. 
where $K_{\mu\nu}$ is the extrinsic curvature \reef{ext}, in which the covariant derivative is replaced by the partial derivative.  Some parts of the above Lagrangian have already been found in \cite{Hosseini:2022vrr} through an analysis of the T-duality of massless closed string fields.
The inclusion of higher-derivative terms of $n_\mu$, such as $\prt K$, presumably involves an infinite number of $F$ terms that may not be written in a compact form like the Lagrangian above. We are not interested in such couplings for this paper.
However, the higher-order couplings that do not involve the vector $n_\mu$ are those that appear in the bulk world-volume Lagrangian, which we anticipate to find at order $\alpha'$. In other words, the boundary world-volume Lagrangian includes all couplings in the bulk Lagrangian as well as other couplings involving the vector field $n_\mu$.

The outline of the paper is as follows:
In Section 2, we study the bulk Lagrangian at order $\alpha'$ up to 8th order of $F$. In Subsection 2.1, we contract all contractions of $F,\nabla F,\nabla\nabla F$, and the second fundamental form $\Omega$ at order $\alpha'$ and remove from them the redundant terms from $A_a$-field redefinition and the Bianchi identity to find 145 independent terms. In Subsection 2.2, we impose the T-duality constraint on these couplings and fix all 145 coupling constants in terms of three parameters. These couplings should be the same as the effective actions that have been found in  \cite{Karimi:2018vaf} up to some total derivative terms and $X^\mu$- and $A_a$-field redefinitions. This fixes two of the three parameters. The remaining parameter is unphysical and can be removed by the most general field redefinitions and integration by parts. The final world-volume Lagrangian  has 49 couplings.
In Section 3, we study the T-duality of the boundary Lagrangian \reef{DBIK}.
In Section 4, we briefly discuss our results. We use the Mathematica package xAct \cite{Nutma:2013zea} for performing the calculations in this paper.

\section{Bulk Lagrangian}

It is known that the bulk Lagrangian \reef{DBI} is consistent with the T-duality transformations (see e.g. \cite{Myers:1999ps}). To prepare our notation for the later sections, we perform a similar calculation here.
We consider a specific background with a circular dimension. That is, the manifold $M^{(D)}$ has the structure $M^{(D)} = M^{(D-1)} \times S^{(1)}$, where $M^{(D-1)}$ is the non-circular part and $S^{(1)}$ is the circular dimension. The coordinates of $M^{(D)}$ are $x^\mu = (x^{\tilde{\mu}}, y)$, where $x^{\tilde{\mu}}$ are the coordinates of $M^{(D-1)}$ and $y$ is the coordinate of $S^{(1)}$. All massless fields are independent of $y$.
There are two reductions of the world-volume action on the circular dimension:

1. When the $D_p$-brane is along the circle, the world-volume indices split as $a=(\tilde{a},y)$, and the spacetime coordinate field splits as $X^\mu(\sigma) = (X^{\tilde{\mu}}(\sigma), y)$. The resulting reduction of the Lagrangian is denoted as ${L}_p^w$.

2. When the $D_{p-1}$-brane is orthogonal to the circle, the world-volume indices do not split, i.e., $a=\tilde{a}$, and the spacetime coordinate field splits as $X^\mu(\sigma) = (X^{\tilde{\mu}}(\sigma), X^y(\sigma))$. The corresponding reduction of the Lagrangian is denoted as ${L}_{p-1}^t$.

 These  two reductions are not identical. However, the transformation of ${L}_p^w$ under the T-duality transformations
\beqa
{A_y} &\rightarrow & {X ^y}\equiv S,\nonumber\\
{A_{\tilde a}} &\rightarrow & {A_{\tilde a}},\nonumber\\
{X ^{\tilde \mu }}&\rightarrow &{X ^{\tilde \mu }},
\labell{a20}
\eeqa
which is called ${L}_{p-1}^{wT}$, should be the same as ${L}_{p-1}^t$. One can easily find that for the following reduction for the  metric:
\beqa
G_{\mu\nu}=
\left(\matrix{
g_{\tmu \tnu} & 0\!\!\!\!\! &\cr
0 &1 \!\!\!\!\!&
}\right)\,,\labell{IG0}
\eeqa
where $g_{\tilde{\mu}\tilde{\nu}}$ is the constant base space metric and 1 is the radius of the circle, the reductions ${L}_p^w$ and ${L}_{p-1}^t$ for the Lagrangian in \reef{DBI} are
 \beqa
 {L}_p^{(0)w}&=&-T_p\sqrt{-\det(\tg_{\ta\tb}+F_{\ta\tb}+\prt_\ta A_y\prt_\tb A_y)}\,,\labell{DBIw}\\
 {L}_{p-1}^{(0)t}&=&-T_{p-1}\sqrt{-\det(\tg_{\ta\tb}+F_{\ta\tb}+\prt_\ta S\prt_\tb S)}\,,\nn
 \eeqa
where in the first line we have used the fact that $\partial_y y = 1$ and $\partial_\ta y = 0$. The pull-back metric in the base space is
\beqa
 \tg_{\ta\tb}&=&\prt_\ta X^\tmu\prt_\tb X^{\tnu} g_{\tmu\tnu}\,.\labell{bg}
\eeqa
The transformation of the first line under the T-duality transformations \reef{a20} becomes identical to the second line above.

The leading-order DBI action \reef{DBI} has an infinite number of field strengths $F$ which appear in a compact form under the square root of the determinant. There are also an infinite number of $F$ terms at each $\alpha'$ correction to the DBI action. Presumably, these higher-order terms cannot be written in a compact form as in the DBI action, which may be a consequence of the fact that at higher orders of $\alpha'$, the T-duality transformations \reef{a20} receive higher-derivative corrections.
However, T-duality still helps us to find the $\alpha'$ corrections to the DBI action \reef{DBI} and the T-duality transformations \reef{a20}. In order to find such corrections, one should first find all independent gauge-invariant couplings at each order of $\alpha'$ with arbitrary coupling constants, and then impose the T-duality transformations, which involve some arbitrary gauge-invariant corrections to \reef{a20}, to determine the coupling constants of the independent terms and the arbitrary parameters of the T-duality transformations.
In fact, one can do the same calculation for the leading-order couplings as well. That is, if one considers all couplings involving $F$ with their indices contracted with the inverse of the pull-back metric $\tG_{ab}$, and then requires them to be invariant under the T-duality transformations \reef{a20}, one finds that all coupling constants are fixed and produce the expansion of the DBI action \reef{DBI}.

At higher orders of $\alpha'$, however, it is non-trivial to find independent couplings in the Lagrangian because there are freedoms of field redefinitions, and also since the couplings involve derivatives of $F$ and $\Omega$ which satisfy their corresponding Bianchi identities, there are freedoms of using Bianchi identities.
In the next section, we will find such independent couplings at order $\alpha'$.

\subsection{Independent Basis}

In this section, we will find the independent covariant and gauge-invariant couplings at order $\alpha'$ which involve at most eight gauge fields and/or the second fundamental form. We use the method introduced in \cite{Metsaev:1987zx,Garousi:2019cdn} to find the independent terms. 
The independent couplings are all gauge-invariant couplings, modulo the field redefinitions and the following Bianchi identities:
\beqa
\nabla_{[a}F_{bc]}&=&0\,,\nn\\
{\Omega _{ab}{}^\mu}{\partial _c}{X^\nu } G_{\mu\nu}&=& 0\,, \labell{Identity}
\eeqa
where the second fundamental form is defined as in \cite{Bachas:1999um}:
\beqa
\Omega_{ab}{}^\mu&=&\nabla_a\prt_b X^\mu\,.\labell{O}
\eeqa
The covariant derivative is constructed from the pull-back metric given in \reef{pull}. Using the second identity in \reef{Identity}, one finds that there is a scheme in which $\partial X$ can appear only through the pull-back metric \reef{pull} and its inverse. For example, the coupling $\nabla\Omega\partial X$ can be written as $-\Omega\Omega$, which can easily be verified by taking the covariant derivative of the above identity.
Hence, we use the scheme in which the couplings involve only the contractions of $F$, $\Omega$, and their covariant derivatives. There are 12 such structures as
\beqa
\bold{L}_p'{}^{(1)}&=&- T_p\alpha'\, \sqrt { - \det (\tilde G_{ab}+F_{ab})}\Big[\Omega^2(1+F^2+F^4+F^6)+(\nabla F)^2(1+F^2+F^4+F^6)\nn\\&&\qquad\qquad\qquad\quad+\nabla\nabla F(F+F^3+F^5+F^7)\Big]\,.\labell{s1}
\eeqa
If one considers all contractions of each structure with the inverse of the pull-back metric, one finds there are 332 couplings in the above Lagrangian. We call the coupling constants of these couplings $a_1', a_2', \cdots, a'_{332}$.
However, they are not all independent couplings. Since we are going to find a Lagrangian that is invariant under T-duality without any residual total derivative terms in the base space, we should not remove the couplings in the above Lagrangian that are related by total derivative terms. It turns out that if one removes the terms in the above Lagrangian that are related by $X^\mu$-field redefinition, then the T-duality is satisfied up to some residual total derivative terms in the base space. Hence, we do not remove the terms in the above Lagrangian that are related by the $X^\mu$-field redefinition either.

We are allowed to use only $A_a$-field redefinition, i.e.:
\beqa
A_a&\rightarrow & A_a+\sqrt{\alpha'}\delta A_a\,.
\eeqa
There are 4 structures $\nabla F (1 + F^2 + F^4 + F^6)$ for constructing the vector $\delta A_a$. If one considers all contractions of these structures with arbitrary parameters, one would find 47 terms in $\delta A_a$.
On the other hand, the leading order Lagrangian \reef{DBI} under the above field redefinitions produces the following couplings at order $\alpha'$:
\beqa
\delta \bold{L}_p^{(0)}=-\frac{T_p\alpha'}{2}\sqrt{-\det(\tG_{ab}+F_{ab})}\Tr\Big[(\tG+F)^{(-1)}\delta F \Big]\,,
\eeqa
where $\delta F_{ab} = \nabla_a \delta A_b - \nabla_b \delta A_a$, and the trace is imposed after expanding $(\tilde{G}+F)^{-1}$.
One should expand the term inside the bracket, i.e. $\Tr\Big[(\tG+F)^{(-1)}\delta F \Big]$ and consider only the terms up to 8th order of $F$. Note that since there is a DBI factor in both the Lagrangian \reef{s1} and in the above equation, we do not expand the DBI factor. Adding these contributions to the Lagrangian \reef{s1}, one finds the result is the same as the Lagrangian \reef{s1} but the coupling constants are changed. If one calls the new coupling constants $a_1, a_2, \cdots, a_{332}$, then one finds:
\beqa
\Delta \bold{L}_p^{(1)}+\delta \bold{L}_p^{(0)}&=&0\,,
\eeqa
where $\Delta \bold{L}_p^{(1)}$ is the same as \reef{s1} with coupling constants $\Delta a_1, \cdots, \Delta a_{332}$ where $\Delta a_i = a_i' - a_i$. Since both terms above have the DBI factor, one can write it as
\beqa
\sqrt{-\det(\tilde{G}_{ab}+F_{ab})}\Delta \mathbf{L}&=&0.
\eeqa
The DBI factor is not zero; hence, the solution of the above equation is the same as the solution of the following equation:
\beqa
\Delta \bold{L}&=&0.
\eeqa
To solve the above equation, one has to impose the first Bianchi identity in \reef{Identity} into it. We impose it by writing the terms that have a derivative of $F_{ab}$ in terms of the potential $A_a$. Then the coefficient of independent terms must be zero, which produces some algebraic equations between the $\Delta a_i$'s and the parameters in the $A_a$-field redefinitions.
We find 145 relations between only $\Delta a_i$'s, which indicates there are 145 independent couplings in \reef{s1}. 
We set the coefficient of all terms in \reef{s1} that have $\nabla\nabla F$ and terms that have $\nabla_a F\nabla^a F$ to zero, and find that the algebraic equations still have 145 relations between only $\Delta a_i$'s, which indicates these terms are allowed to be zero. We also add some other terms to be zero such that we find 145 relations $\Delta a_i=0$. The independent couplings in the particular scheme that we have chosen are the following:
\beqa
\bold{L}_p^{(1)}&\!\!\!\!\!=\!\!\!\!\!&-T_p\sqrt{-\det(\tG_{ab}+F_{ab})}\Big[  a_{1}   \Omega _{ab\mu  } \Omega ^{ab\mu  }\! + \!  
   a_{2}  
    F^{ab} F^{cd} \Omega _{ac}{}^{\mu  } \Omega _{bd\mu  }\! + \! 
    a_{3}  
    \Omega ^{a}{}_{a}{}^{\mu  } \Omega ^{b}{}_{b\mu  } \!+ \!  
   a_{4}  
    F_{a}{}^{c} F^{ab} \Omega _{b}{}^{d\mu  } \Omega _{cd\mu  } \nn\\&\!\!\!\!\!\!\!&
+   a_{5}  
    F_{a}{}^{c} F^{ab} F_{d}{}^{f} F^{de} \Omega _{be}{}^{\mu  } 
\Omega _{cf\mu  } +   a_{6}  
    F_{ab} F^{ab} \Omega _{cd\mu  } \Omega ^{cd\mu  } +   
   a_{7}  
    F_{a}{}^{c} F^{ab} F_{b}{}^{d} F^{ef} \Omega _{ce}{}^{\mu  } 
\Omega _{df\mu  }\nn\\&\!\!\!\!\!\!\!& +   a_{8}  
    F_{ab} F^{ab} F^{cd} F^{ef} \Omega _{ce}{}^{\mu  } \Omega 
_{df\mu  } +   a_{9}  
    F_{a}{}^{c} F^{ab} F_{b}{}^{d} F_{e}{}^{h} F^{ef} F_{f}{}^{i} 
\Omega _{ch}{}^{\mu  } \Omega _{di\mu  } +   a_{10}  
    F_{a}{}^{c} F^{ab} \Omega _{bc}{}^{\mu  } \Omega 
^{d}{}_{d\mu  }\nn\\&\!\!\!\!\!\!\!& +   a_{11}  
    F_{ab} F^{ab} \Omega ^{c}{}_{c}{}^{\mu  } \Omega 
^{d}{}_{d\mu  } +   a_{12}  
    F_{a}{}^{c} F^{ab} F_{b}{}^{d} F_{c}{}^{e} \Omega _{d}{}^{f\mu 
 } \Omega _{ef\mu  } +   a_{13}  
    F_{ab} F^{ab} F_{c}{}^{e} F^{cd} \Omega _{d}{}^{f\mu  } 
\Omega _{ef\mu  } \nn\\&\!\!\!\!\!\!\!&+   a_{14}  
    F_{a}{}^{c} F^{ab} F_{b}{}^{d} F_{c}{}^{e} F_{f}{}^{i} F^{fh} 
\Omega _{dh}{}^{\mu  } \Omega _{ei\mu  } +   a_{15}  
    F_{ab} F^{ab} F_{c}{}^{e} F^{cd} F_{f}{}^{i} F^{fh} \Omega 
_{dh}{}^{\mu  } \Omega _{ei\mu  }\nn\\&\!\!\!\!\!\!\!& +\!   a_{16}  
    F_{a}{}^{c} F^{ab} F_{b}{}^{d} F_{cd} \Omega _{ef\mu  } 
\Omega ^{ef\mu  }\! +\!   a_{17}  
    F_{ab} F^{ab} F_{cd} F^{cd} \Omega _{ef\mu  } \Omega ^{ef\mu 
 } \!+\!   a_{18}  
    F_{a}{}^{c} F^{ab} F_{b}{}^{d} F_{c}{}^{e} F_{d}{}^{f} F^{hi} 
\Omega _{eh}{}^{\mu  } \Omega _{fi\mu  } \nn\\&\!\!\!\!\!\!\!&+   a_{19}  
    F_{ab} F^{ab} F_{c}{}^{e} F^{cd} F_{d}{}^{f} F^{hi} \Omega 
_{eh}{}^{\mu  } \Omega _{fi\mu  } +   a_{20}  
    F_{a}{}^{c} F^{ab} F_{b}{}^{d} F_{cd} F^{ef} F^{hi} \Omega 
_{eh}{}^{\mu  } \Omega _{fi\mu  } \nn\\&\!\!\!\!\!\!\!&+ \!  a_{21}  
    F_{ab} F^{ab} F_{cd} F^{cd} F^{ef} F^{hi} \Omega _{eh}{}^{\mu  
} \Omega _{fi\mu  } \!+ \!  a_{22}  
    F_{a}{}^{c} F^{ab} F_{b}{}^{d} F_{c}{}^{e} \Omega _{de}{}^{\mu 
 } \Omega ^{f}{}_{f\mu  }\! + \!  a_{23}  
    F_{ab} F^{ab} F_{c}{}^{e} F^{cd} \Omega _{de}{}^{\mu  } 
\Omega ^{f}{}_{f\mu  } \nn\\&\!\!\!\!\!\!\!&+\!   a_{24}  
    F_{a}{}^{c} F^{ab} F_{b}{}^{d} F_{cd} \Omega ^{e}{}_{e}{}^{\mu 
 } \Omega ^{f}{}_{f\mu  }\! + \!  a_{25}  
    F_{ab} F^{ab} F_{cd} F^{cd} \Omega ^{e}{}_{e}{}^{\mu  } 
\Omega ^{f}{}_{f\mu  } \!+ \!  a_{26}  
    F_{a}{}^{c} F^{ab} F_{b}{}^{d} F_{c}{}^{e} F_{f}{}^{i} F^{fh} 
\Omega _{de}{}^{\mu  } \Omega _{hi\mu  } \nn\\&\!\!\!\!\!\!\!&+   a_{27}  
    F_{ab} F^{ab} F_{c}{}^{e} F^{cd} F_{f}{}^{i} F^{fh} \Omega 
_{de}{}^{\mu  } \Omega _{hi\mu  } +   a_{28}  
    F_{a}{}^{c} F^{ab} F_{b}{}^{d} F_{c}{}^{e} F_{d}{}^{f} 
F_{e}{}^{h} \Omega _{f}{}^{i\mu  } \Omega _{hi\mu  } \nn\\&\!\!\!\!\!\!\!&+   
   a_{29}  
    F_{ab} F^{ab} F_{c}{}^{e} F^{cd} F_{d}{}^{f} F_{e}{}^{h} \Omega 
_{f}{}^{i\mu  } \Omega _{hi\mu  } +   a_{30}  
    F_{a}{}^{c} F^{ab} F_{b}{}^{d} F_{cd} F_{e}{}^{h} F^{ef} \Omega 
_{f}{}^{i\mu  } \Omega _{hi\mu  }\nn\\&\!\!\!\!\!\!\!& +   a_{31}  
    F_{ab} F^{ab} F_{cd} F^{cd} F_{e}{}^{h} F^{ef} \Omega _{f}{}^{i
\mu  } \Omega _{hi\mu  } +   a_{32}  
    F_{a}{}^{c} F^{ab} F_{b}{}^{d} F_{c}{}^{e} F_{d}{}^{f} F_{ef} 
\Omega _{hi\mu  } \Omega ^{hi\mu  }\nn\\&\!\!\!\!\!\!\!& +   a_{33}  
    F_{ab} F^{ab} F_{c}{}^{e} F^{cd} F_{d}{}^{f} F_{ef} \Omega _{hi
\mu  } \Omega ^{hi\mu  } +   a_{34}  
    F_{ab} F^{ab} F_{cd} F^{cd} F_{ef} F^{ef} \Omega _{hi\mu  } 
\Omega ^{hi\mu  } \nn\\&\!\!\!\!\!\!\!&+   a_{35}  
    F_{a}{}^{c} F^{ab} F_{b}{}^{d} F_{c}{}^{e} F_{d}{}^{f} 
F_{e}{}^{h} \Omega _{fh}{}^{\mu  } \Omega ^{i}{}_{i\mu  } +   
   a_{36}  
    F_{ab} F^{ab} F_{c}{}^{e} F^{cd} F_{d}{}^{f} F_{e}{}^{h} \Omega 
_{fh}{}^{\mu  } \Omega ^{i}{}_{i\mu  } \nn\\&\!\!\!\!\!\!\!&+   a_{37}  
    F_{a}{}^{c} F^{ab} F_{b}{}^{d} F_{cd} F_{e}{}^{h} F^{ef} \Omega 
_{fh}{}^{\mu  } \Omega ^{i}{}_{i\mu  } +   a_{38}  
    F_{ab} F^{ab} F_{cd} F^{cd} F_{e}{}^{h} F^{ef} \Omega _{fh}{}^{
\mu  } \Omega ^{i}{}_{i\mu  } \nn\\&\!\!\!\!\!\!\!&+   a_{39}  
    F_{a}{}^{c} F^{ab} F_{b}{}^{d} F_{c}{}^{e} F_{d}{}^{f} F_{ef} 
\Omega ^{h}{}_{h}{}^{\mu  } \Omega ^{i}{}_{i\mu  } +   
   a_{40}  
    F_{ab} F^{ab} F_{c}{}^{e} F^{cd} F_{d}{}^{f} F_{ef} \Omega 
^{h}{}_{h}{}^{\mu  } \Omega ^{i}{}_{i\mu  }\nn\\&\!\!\!\!\!\!\!& +   a_{41}  
    F_{ab} F^{ab} F_{cd} F^{cd} F_{ef} F^{ef} \Omega ^{h}{}_{h}{}^{
\mu  } \Omega ^{i}{}_{i\mu  } +   a_{42}  
    \nabla _{a}F^{ab} \nabla _{c}F_{b}{}^{c} +   a_{43}  
    \nabla _{b}F_{ac} \nabla ^{c}F^{ab} \nn\\&\!\!\!\!\!\!\!&+   a_{44}  
    F^{ab} F^{cd} \nabla _{c}F_{a}{}^{e} \nabla _{d}F_{be} +   
   a_{45}  
    F_{a}{}^{c} F^{ab} \nabla _{d}F_{b}{}^{d} \nabla 
_{e}F_{c}{}^{e} +   a_{46}  
    F^{ab} F^{cd} \nabla _{c}F_{ab} \nabla _{e}F_{d}{}^{e} \nn\\&\!\!\!\!\!\!\!&+   
   a_{47}  
    F_{a}{}^{c} F^{ab} \nabla _{c}F_{b}{}^{d} \nabla 
_{e}F_{d}{}^{e} +   a_{48}  
    F_{ab} F^{ab} \nabla _{c}F^{cd} \nabla _{e}F_{d}{}^{e} +   
   a_{49}  
    F^{ab} F^{cd} \nabla _{d}F_{ce} \nabla ^{e}F_{ab} \nn\\&\!\!\!\!\!\!\!&+   
   a_{50}  
    F^{ab} F^{cd} \nabla _{d}F_{be} \nabla ^{e}F_{ac} +   
   a_{51}  
    F_{a}{}^{c} F^{ab} \nabla _{c}F_{de} \nabla ^{e}F_{b}{}^{d} + 
  a_{52}  
    F_{a}{}^{c} F^{ab} \nabla _{d}F_{ce} \nabla ^{e}F_{b}{}^{d}\nn\\&\!\!\!\!\!\!\!& + 
  a_{53}  
    F_{ab} F^{ab} \nabla _{d}F_{ce} \nabla ^{e}F^{cd} +   
   a_{54}  
    F_{a}{}^{c} F^{ab} F_{b}{}^{d} F^{ef} \nabla _{e}F_{c}{}^{h} 
\nabla _{f}F_{dh} +   a_{55}  
    F_{a}{}^{c} F^{ab} F_{d}{}^{f} F^{de} \nabla _{c}F_{b}{}^{h} 
\nabla _{f}F_{eh}\nn\\&\!\!\!\!\!\!\!& +   a_{56}  
    F_{a}{}^{c} F^{ab} F^{de} F^{fh} \nabla _{f}F_{bd} \nabla 
_{h}F_{ce} +   a_{57}  
    F_{a}{}^{c} F^{ab} F^{de} F^{fh} \nabla _{e}F_{bd} \nabla 
_{h}F_{cf} \nn\\&\!\!\!\!\!\!\!&+   a_{58}  
    F_{a}{}^{c} F^{ab} F_{b}{}^{d} F^{ef} \nabla _{f}F_{ce} \nabla 
_{h}F_{d}{}^{h} +   a_{59}  
    F_{a}{}^{c} F^{ab} F^{de} F^{fh} \nabla _{c}F_{bd} \nabla 
_{h}F_{ef}\nn\\&\!\!\!\!\!\!\!& +   a_{60}  
    F_{a}{}^{c} F^{ab} F_{b}{}^{d} F_{c}{}^{e} \nabla 
_{f}F_{d}{}^{f} \nabla _{h}F_{e}{}^{h} +   a_{61}  
    F_{ab} F^{ab} F_{c}{}^{e} F^{cd} \nabla _{f}F_{d}{}^{f} \nabla 
_{h}F_{e}{}^{h}\nn\\&\!\!\!\!\!\!\!& +   a_{62}  
    F_{a}{}^{c} F^{ab} F_{d}{}^{f} F^{de} \nabla _{c}F_{be} \nabla 
_{h}F_{f}{}^{h} +   a_{63}  
    F_{a}{}^{c} F^{ab} F_{b}{}^{d} F^{ef} \nabla _{e}F_{cd} \nabla 
_{h}F_{f}{}^{h} \nn\\&\!\!\!\!\!\!\!&+   a_{64}  
    F_{ab} F^{ab} F^{cd} F^{ef} \nabla _{e}F_{cd} \nabla 
_{h}F_{f}{}^{h} +   a_{65}  
    F_{a}{}^{c} F^{ab} F_{b}{}^{d} F_{c}{}^{e} \nabla 
_{e}F_{d}{}^{f} \nabla _{h}F_{f}{}^{h}\nn\\&\!\!\!\!\!\!\!& +   a_{66}  
    F_{ab} F^{ab} F_{c}{}^{e} F^{cd} \nabla _{e}F_{d}{}^{f} \nabla 
_{h}F_{f}{}^{h} +   a_{67}  
    F_{a}{}^{c} F^{ab} F_{b}{}^{d} F_{cd} \nabla _{e}F^{ef} \nabla 
_{h}F_{f}{}^{h}\nn\\&\!\!\!\!\!\!\!& +   a_{68}  
    F_{ab} F^{ab} F_{cd} F^{cd} \nabla _{e}F^{ef} \nabla 
_{h}F_{f}{}^{h} +   a_{69}  
    F_{a}{}^{c} F^{ab} F_{d}{}^{f} F^{de} \nabla _{f}F_{ch} \nabla 
^{h}F_{be} \nn\\&\!\!\!\!\!\!\!&+   a_{70}  
    F_{a}{}^{c} F^{ab} F_{b}{}^{d} F^{ef} \nabla _{f}F_{eh} \nabla 
^{h}F_{cd} +   a_{71}  
    F_{ab} F^{ab} F^{cd} F^{ef} \nabla _{f}F_{eh} \nabla 
^{h}F_{cd} \nn\\&\!\!\!\!\!\!\!&+   a_{72}  
    F_{a}{}^{c} F^{ab} F_{b}{}^{d} F^{ef} \nabla _{d}F_{fh} \nabla 
^{h}F_{ce} +   a_{73}  
    F_{a}{}^{c} F^{ab} F_{b}{}^{d} F^{ef} \nabla _{f}F_{dh} \nabla 
^{h}F_{ce} \nn\\&\!\!\!\!\!\!\!&+   a_{74}  
    F_{ab} F^{ab} F^{cd} F^{ef} \nabla _{f}F_{dh} \nabla 
^{h}F_{ce} +   a_{75}  
    F_{a}{}^{c} F^{ab} F_{b}{}^{d} F_{c}{}^{e} \nabla _{e}F_{fh} 
\nabla ^{h}F_{d}{}^{f} \nn\\&\!\!\!\!\!\!\!&+   a_{76}  
    F_{ab} F^{ab} F_{c}{}^{e} F^{cd} \nabla _{e}F_{fh} \nabla 
^{h}F_{d}{}^{f} +   a_{77}  
    F_{a}{}^{c} F^{ab} F_{b}{}^{d} F_{c}{}^{e} \nabla _{f}F_{eh} 
\nabla ^{h}F_{d}{}^{f} \nn\\&\!\!\!\!\!\!\!&+   a_{78}  
    F_{ab} F^{ab} F_{c}{}^{e} F^{cd} \nabla _{f}F_{eh} \nabla 
^{h}F_{d}{}^{f} +   a_{79}  
    F_{a}{}^{c} F^{ab} F_{b}{}^{d} F_{cd} \nabla _{f}F_{eh} \nabla 
^{h}F^{ef}\nn\\&\!\!\!\!\!\!\!& +   a_{80}  
    F_{ab} F^{ab} F_{cd} F^{cd} \nabla _{f}F_{eh} \nabla 
^{h}F^{ef} +   a_{81}  
    F_{a}{}^{c} F^{ab} F_{b}{}^{d} F_{e}{}^{h} F^{ef} F_{f}{}^{i} 
\nabla _{h}F_{c}{}^{j} \nabla _{i}F_{dj} \nn\\&\!\!\!\!\!\!\!&+   a_{82}  
    F_{a}{}^{c} F^{ab} F_{b}{}^{d} F_{c}{}^{e} F_{f}{}^{i} F^{fh} 
\nabla _{h}F_{d}{}^{j} \nabla _{i}F_{ej} +   a_{83}  
    F_{a}{}^{c} F^{ab} F_{b}{}^{d} F_{c}{}^{e} F_{d}{}^{f} F^{hi} 
\nabla _{h}F_{e}{}^{j} \nabla _{i}F_{fj}\nn\\&\!\!\!\!\!\!\!& +   a_{84}  
    F_{ab} F^{ab} F_{c}{}^{e} F^{cd} F_{d}{}^{f} F^{hi} \nabla 
_{h}F_{e}{}^{j} \nabla _{i}F_{fj} +   a_{85}  
    F_{a}{}^{c} F^{ab} F_{b}{}^{d} F_{cd} F^{ef} F^{hi} \nabla 
_{h}F_{e}{}^{j} \nabla _{i}F_{fj} \nn\\&\!\!\!\!\!\!\!&+   a_{86}  
    F_{a}{}^{c} F^{ab} F_{b}{}^{d} F_{c}{}^{e} F_{f}{}^{i} F^{fh} 
\nabla _{e}F_{d}{}^{j} \nabla _{i}F_{hj} +   a_{87}  
    F_{ab} F^{ab} F_{c}{}^{e} F^{cd} F_{f}{}^{i} F^{fh} \nabla 
_{e}F_{d}{}^{j} \nabla _{i}F_{hj}\nn\\&\!\!\!\!\!\!\!& +   a_{88}  
    F_{a}{}^{c} F^{ab} F_{d}{}^{f} F^{de} F_{h}{}^{j} F^{hi} \nabla 
_{i}F_{be} \nabla _{j}F_{cf} +   a_{89}  
    F_{a}{}^{c} F^{ab} F_{b}{}^{d} F_{e}{}^{h} F^{ef} F^{ij} \nabla 
_{i}F_{cf} \nabla _{j}F_{dh} \nn\\&\!\!\!\!\!\!\!&+   a_{90}  
    F_{a}{}^{c} F^{ab} F_{b}{}^{d} F_{e}{}^{h} F^{ef} F^{ij} \nabla 
_{h}F_{cf} \nabla _{j}F_{di} +   a_{91}  
    F_{a}{}^{c} F^{ab} F_{b}{}^{d} F_{c}{}^{e} F^{fh} F^{ij} \nabla 
_{i}F_{df} \nabla _{j}F_{eh} \nn\\&\!\!\!\!\!\!\!&+   a_{92}  
    F_{ab} F^{ab} F_{c}{}^{e} F^{cd} F^{fh} F^{ij} \nabla 
_{i}F_{df} \nabla _{j}F_{eh} +   a_{93}  
    F_{a}{}^{c} F^{ab} F_{b}{}^{d} F_{c}{}^{e} F^{fh} F^{ij} \nabla 
_{h}F_{df} \nabla _{j}F_{ei}\nn\\&\!\!\!\!\!\!\!& +   a_{94}  
    F_{ab} F^{ab} F_{c}{}^{e} F^{cd} F^{fh} F^{ij} \nabla 
_{h}F_{df} \nabla _{j}F_{ei} +   a_{95}  
    F_{a}{}^{c} F^{ab} F_{b}{}^{d} F_{c}{}^{e} F_{f}{}^{i} F^{fh} 
\nabla _{i}F_{dh} \nabla _{j}F_{e}{}^{j} \nn\\&\!\!\!\!\!\!\!&+   a_{96}  
    F_{a}{}^{c} F^{ab} F_{d}{}^{f} F^{de} F_{h}{}^{j} F^{hi} \nabla 
_{c}F_{be} \nabla _{j}F_{fi} +   a_{97}  
    F_{a}{}^{c} F^{ab} F_{b}{}^{d} F_{c}{}^{e} F_{d}{}^{f} F^{hi} 
\nabla _{i}F_{eh} \nabla _{j}F_{f}{}^{j}\nn\\&\!\!\!\!\!\!\!& +   a_{98}  
    F_{ab} F^{ab} F_{c}{}^{e} F^{cd} F_{d}{}^{f} F^{hi} \nabla 
_{i}F_{eh} \nabla _{j}F_{f}{}^{j} +   a_{99}  
    F_{a}{}^{c} F^{ab} F_{b}{}^{d} F_{c}{}^{e} F^{fh} F^{ij} \nabla 
_{e}F_{df} \nabla _{j}F_{hi} \nn\\&\!\!\!\!\!\!\!&+   a_{100}  
    F_{ab} F^{ab} F_{c}{}^{e} F^{cd} F^{fh} F^{ij} \nabla 
_{e}F_{df} \nabla _{j}F_{hi} +   a_{101}  
    F_{a}{}^{c} F^{ab} F_{b}{}^{d} F_{e}{}^{h} F^{ef} F^{ij} \nabla 
_{f}F_{cd} \nabla _{j}F_{hi} \nn\\&\!\!\!\!\!\!\!&+   a_{102}  
    F_{a}{}^{c} F^{ab} F_{b}{}^{d} F_{c}{}^{e} F_{d}{}^{f} 
F_{e}{}^{h} \nabla _{i}F_{f}{}^{i} \nabla _{j}F_{h}{}^{j} +   
   a_{103}  
    F_{ab} F^{ab} F_{c}{}^{e} F^{cd} F_{d}{}^{f} F_{e}{}^{h} \nabla 
_{i}F_{f}{}^{i} \nabla _{j}F_{h}{}^{j} \nn\\&\!\!\!\!\!\!\!&+   a_{104}  
    F_{a}{}^{c} F^{ab} F_{b}{}^{d} F_{cd} F_{e}{}^{h} F^{ef} \nabla 
_{i}F_{f}{}^{i} \nabla _{j}F_{h}{}^{j} +   a_{105}  
    F_{ab} F^{ab} F_{cd} F^{cd} F_{e}{}^{h} F^{ef} \nabla 
_{i}F_{f}{}^{i} \nabla _{j}F_{h}{}^{j} \nn\\&\!\!\!\!\!\!\!&+   a_{106}  
    F_{a}{}^{c} F^{ab} F_{b}{}^{d} F_{c}{}^{e} F_{f}{}^{i} F^{fh} 
\nabla _{e}F_{dh} \nabla _{j}F_{i}{}^{j} +   a_{107}  
    F_{ab} F^{ab} F_{c}{}^{e} F^{cd} F_{f}{}^{i} F^{fh} \nabla 
_{e}F_{dh} \nabla _{j}F_{i}{}^{j} \nn\\&\!\!\!\!\!\!\!&+   a_{108}  
    F_{a}{}^{c} F^{ab} F_{b}{}^{d} F_{e}{}^{h} F^{ef} F_{f}{}^{i} 
\nabla _{h}F_{cd} \nabla _{j}F_{i}{}^{j} +   a_{109}  
    F_{a}{}^{c} F^{ab} F_{b}{}^{d} F_{c}{}^{e} F_{d}{}^{f} F^{hi} 
\nabla _{h}F_{ef} \nabla _{j}F_{i}{}^{j} \nn\\&\!\!\!\!\!\!\!&+   a_{110}  
    F_{ab} F^{ab} F_{c}{}^{e} F^{cd} F_{d}{}^{f} F^{hi} \nabla 
_{h}F_{ef} \nabla _{j}F_{i}{}^{j} +   a_{111}  
    F_{a}{}^{c} F^{ab} F_{b}{}^{d} F_{cd} F^{ef} F^{hi} \nabla 
_{h}F_{ef} \nabla _{j}F_{i}{}^{j} \nn\\&\!\!\!\!\!\!\!&+   a_{112}  
    F_{ab} F^{ab} F_{cd} F^{cd} F^{ef} F^{hi} \nabla _{h}F_{ef} 
\nabla _{j}F_{i}{}^{j} +   a_{113}  
    F_{a}{}^{c} F^{ab} F_{b}{}^{d} F_{c}{}^{e} F_{d}{}^{f} 
F_{e}{}^{h} \nabla _{h}F_{f}{}^{i} \nabla _{j}F_{i}{}^{j}\nn\\&\!\!\!\!\!\!\!& +   
   a_{114}  
    F_{ab} F^{ab} F_{c}{}^{e} F^{cd} F_{d}{}^{f} F_{e}{}^{h} \nabla 
_{h}F_{f}{}^{i} \nabla _{j}F_{i}{}^{j} +   a_{115}  
    F_{a}{}^{c} F^{ab} F_{b}{}^{d} F_{cd} F_{e}{}^{h} F^{ef} \nabla 
_{h}F_{f}{}^{i} \nabla _{j}F_{i}{}^{j} \nn\\&\!\!\!\!\!\!\!&+   a_{116}  
    F_{ab} F^{ab} F_{cd} F^{cd} F_{e}{}^{h} F^{ef} \nabla 
_{h}F_{f}{}^{i} \nabla _{j}F_{i}{}^{j} +   a_{117}  
    F_{a}{}^{c} F^{ab} F_{b}{}^{d} F_{c}{}^{e} F_{d}{}^{f} F_{ef} 
\nabla _{h}F^{hi} \nabla _{j}F_{i}{}^{j} \nn\\&\!\!\!\!\!\!\!&+   a_{118}  
    F_{ab} F^{ab} F_{c}{}^{e} F^{cd} F_{d}{}^{f} F_{ef} \nabla 
_{h}F^{hi} \nabla _{j}F_{i}{}^{j} +   a_{119}  
    F_{ab} F^{ab} F_{cd} F^{cd} F_{ef} F^{ef} \nabla _{h}F^{hi} 
\nabla _{j}F_{i}{}^{j} \nn\\&\!\!\!\!\!\!\!&+   a_{120}  
    F_{a}{}^{c} F^{ab} F_{b}{}^{d} F_{e}{}^{h} F^{ef} F_{f}{}^{i} 
\nabla _{i}F_{hj} \nabla ^{j}F_{cd} +   a_{121}  
    F_{a}{}^{c} F^{ab} F_{b}{}^{d} F_{e}{}^{h} F^{ef} F_{f}{}^{i} 
\nabla _{i}F_{dj} \nabla ^{j}F_{ch} \nn\\&\!\!\!\!\!\!\!&+   a_{122}  
    F_{a}{}^{c} F^{ab} F_{b}{}^{d} F_{c}{}^{e} F_{f}{}^{i} F^{fh} 
\nabla _{e}F_{ij} \nabla ^{j}F_{dh} +   a_{123}  
    F_{a}{}^{c} F^{ab} F_{b}{}^{d} F_{c}{}^{e} F_{f}{}^{i} F^{fh} 
\nabla _{i}F_{ej} \nabla ^{j}F_{dh} \nn\\&\!\!\!\!\!\!\!&+   a_{124}  
    F_{ab} F^{ab} F_{c}{}^{e} F^{cd} F_{f}{}^{i} F^{fh} \nabla 
_{i}F_{ej} \nabla ^{j}F_{dh} +   a_{125}  
    F_{a}{}^{c} F^{ab} F_{b}{}^{d} F_{c}{}^{e} F_{d}{}^{f} F^{hi} 
\nabla _{i}F_{hj} \nabla ^{j}F_{ef} \nn\\&\!\!\!\!\!\!\!&+   a_{126}  
    F_{ab} F^{ab} F_{c}{}^{e} F^{cd} F_{d}{}^{f} F^{hi} \nabla 
_{i}F_{hj} \nabla ^{j}F_{ef} +   a_{127}  
    F_{a}{}^{c} F^{ab} F_{b}{}^{d} F_{cd} F^{ef} F^{hi} \nabla 
_{i}F_{hj} \nabla ^{j}F_{ef} \nn\\&\!\!\!\!\!\!\!&+   a_{128}  
    F_{ab} F^{ab} F_{cd} F^{cd} F^{ef} F^{hi} \nabla _{i}F_{hj} 
\nabla ^{j}F_{ef} +   a_{129}  
    F_{a}{}^{c} F^{ab} F_{b}{}^{d} F_{c}{}^{e} F_{d}{}^{f} F^{hi} 
\nabla _{f}F_{ij} \nabla ^{j}F_{eh} \nn\\&\!\!\!\!\!\!\!&+   a_{130}  
    F_{ab} F^{ab} F_{c}{}^{e} F^{cd} F_{d}{}^{f} F^{hi} \nabla 
_{f}F_{ij} \nabla ^{j}F_{eh} +   a_{131}  
    F_{a}{}^{c} F^{ab} F_{b}{}^{d} F_{c}{}^{e} F_{d}{}^{f} F^{hi} 
\nabla _{i}F_{fj} \nabla ^{j}F_{eh} \nn\\&\!\!\!\!\!\!\!&+   a_{132}  
    F_{ab} F^{ab} F_{c}{}^{e} F^{cd} F_{d}{}^{f} F^{hi} \nabla 
_{i}F_{fj} \nabla ^{j}F_{eh} +   a_{133}  
    F_{a}{}^{c} F^{ab} F_{b}{}^{d} F_{cd} F^{ef} F^{hi} \nabla 
_{i}F_{fj} \nabla ^{j}F_{eh} \nn\\&\!\!\!\!\!\!\!&+   a_{134}  
    F_{ab} F^{ab} F_{cd} F^{cd} F^{ef} F^{hi} \nabla _{i}F_{fj} 
\nabla ^{j}F_{eh} +   a_{135}  
    F_{a}{}^{c} F^{ab} F_{b}{}^{d} F_{c}{}^{e} F_{d}{}^{f} 
F_{e}{}^{h} \nabla _{h}F_{ij} \nabla ^{j}F_{f}{}^{i} \nn\\&\!\!\!\!\!\!\!&+   
   a_{136}  
    F_{ab} F^{ab} F_{c}{}^{e} F^{cd} F_{d}{}^{f} F_{e}{}^{h} \nabla 
_{h}F_{ij} \nabla ^{j}F_{f}{}^{i} +   a_{137}  
    F_{a}{}^{c} F^{ab} F_{b}{}^{d} F_{cd} F_{e}{}^{h} F^{ef} \nabla 
_{h}F_{ij} \nabla ^{j}F_{f}{}^{i} \nn\\&\!\!\!\!\!\!\!&+   a_{138}  
    F_{ab} F^{ab} F_{cd} F^{cd} F_{e}{}^{h} F^{ef} \nabla 
_{h}F_{ij} \nabla ^{j}F_{f}{}^{i} +   a_{139}  
    F_{a}{}^{c} F^{ab} F_{b}{}^{d} F_{c}{}^{e} F_{d}{}^{f} 
F_{e}{}^{h} \nabla _{i}F_{hj} \nabla ^{j}F_{f}{}^{i} \nn\\&\!\!\!\!\!\!\!&+   
   a_{140}  
    F_{ab} F^{ab} F_{c}{}^{e} F^{cd} F_{d}{}^{f} F_{e}{}^{h} \nabla 
_{i}F_{hj} \nabla ^{j}F_{f}{}^{i} +   a_{141}  
    F_{a}{}^{c} F^{ab} F_{b}{}^{d} F_{cd} F_{e}{}^{h} F^{ef} \nabla 
_{i}F_{hj} \nabla ^{j}F_{f}{}^{i} \nn\\&\!\!\!\!\!\!\!&+   a_{142}  
    F_{ab} F^{ab} F_{cd} F^{cd} F_{e}{}^{h} F^{ef} \nabla 
_{i}F_{hj} \nabla ^{j}F_{f}{}^{i} +   a_{143}  
    F_{a}{}^{c} F^{ab} F_{b}{}^{d} F_{c}{}^{e} F_{d}{}^{f} F_{ef} 
\nabla _{i}F_{hj} \nabla ^{j}F^{hi} \nn\\&\!\!\!\!\!\!\!&+   a_{144}  
    F_{ab} F^{ab} F_{c}{}^{e} F^{cd} F_{d}{}^{f} F_{ef} \nabla 
_{i}F_{hj} \nabla ^{j}F^{hi} +   a_{145}  
    F_{ab} F^{ab} F_{cd} F^{cd} F_{ef} F^{ef} \nabla _{i}F_{hj} 
\nabla ^{j}F^{hi}\Big],\labell{Lp}
\eeqa
where $a_1,\cdots, a_{145}$ are 145 background-independent coupling constants that should be fixed by the T-duality symmetry when the spacetime has one circle or by the S-matrix method when the spacetime background is flat.

If one removes the terms in \reef{s1} that are related by total derivative terms and by $X^\mu$-field redefinitions, 81 independent terms are obtained, as found in \cite{Karimi:2018vaf}. The T-duality is satisfied for these 81 couplings up to some anomalous total derivative terms in the base space. The T-duality would fix these 81 couplings up to two parameters, and the two parameters are also fixed in \cite{Karimi:2018vaf} by comparing the resulting four-field couplings with the disk-level four-point S-matrix element. These results appear  in Appendix A. In the next subsection, we will impose T-duality on the basis in \reef{Lp} to fix its coupling constants.

\subsection{T-Duality Constraint on the Basis}

To study the T-duality constraint on the coupling constants in \reef{Lp}, one should first peform the circular reduction in the two cases as in \reef{DBIw}.  In the first case, the reduction of pull-back metric and $F$ are
\beqa
\tG_{ab}=
\left(\matrix{
\tg_{\ta \tb} & 0\!\!\!\!\! &\cr
0 &1 \!\!\!\!\!&
}\right),\,\,F_{ab}=
\left(\matrix{
F_{\ta \tb} & \prt_\ta A_y\!\!\!\!\! &\cr
-\prt_\tb A_y &0 \!\!\!\!\!&
}\right)\,.
\eeqa
Then, the reduction of the DBI factor in  \reef{Lp} is the same as the first term in  \reef{DBIw}. Using the fact that in the dimensional reduction, one assumes fields are independent of the $y$ coordinate, and using the above reduction of the pull-back metric, one finds that the covariant derivative constructed from the pull-back metric $\tilde{G}_{ab}$ reduces as follows:
\beqa
\nabla_y(\cdots)=0\,,&& \nabla_\ta(\cdots)=\tilde{\nabla}_\ta(\cdots)\,,
\eeqa
where $\tilde{\nabla}$ is covariant derivative construncted from the base space pull-back metric $\tg_{\ta\tb}$. In particular, the only non-zero component of the second fundamental form is $\Omega_{\ta\tb}{}^{\tmu}=\tilde{\nabla}_\ta\prt_\tb{}X^\tmu\equiv \tilde{\Omega}_{\ta\tb}{}^{\tmu}$. In this way, one can easily calculate ${L}_p^{(1)w}$. Then using the transformation \reef{a20}, one can calculate  ${L}_{p-1}^{(1)wT}$.

In the second case where $D_{p-1}$-brane is orthogonal to the circle, there is no world-volume index $y$. So the $F$ and pull-back metric has only world-volume indices. The pull-back metric is
\beqa
{{\tilde G}_{\ta\tb}} &=& {{\tilde g}_{\tilde a\tilde b}} + {\partial _{\tilde a}}{S}{\partial _{\tilde b}}{S}\,.\labell{pull2}
\eeqa
Then the reduction of the DBI factor in  \reef{Lp} is the same as the second term in  \reef{DBIw}. The above reduction of the pull-back metric allows the covariant derivative $\nabla_\ta$ constructed from the pull-back metric $\tilde{G}_{\ta\tb}$, which involves the inverse of this metric in the Christoffel connection, to have an expansion in terms of $\partial_{\tilde a} S\partial_{\tilde b} S$. This allows the reduction of $\Omega$ and $\nabla F$ to be
\beqa
\Omega_{\ta\tb}{}^y&=&\tilde{\Omega}_{\ta\tb}{}^y\Big[1+\tg^{\tc\td}\prt_\tc S\prt_\td S\Big]^{-1}\,,\nn\\
\Omega_{\ta\tb}{}^{\tmu}&=&\tilde{\Omega}_{\ta\tb}{}^{\tmu}-\tilde{\Omega}_{\ta\tb}{}^y\prt_\tc S\prt_\td X^\tmu\tG^{\tc\td}\,,\nn\\
\nabla_\ta F_{\tb\tc}&=&\tilde{\nabla}_\ta F_{\tb\tc}+\tilde{\Omega}_{\ta\tb}{}^y\prt_\td SF_{\tc\te}\tG^{\td\te}-\tilde{\Omega}_{\ta\tc}{}^y\prt_\td SF_{\tb\te}\tG^{\td\te}\,,
\labell{a22}
\eeqa
where  $\tilde{\Omega}_{\ta\tb}{}^{y}=\tilde{\nabla}_\ta\prt_\tb{}S$,  and $\tG^{\ta\tb}$ is inverse of the pull-back metric \reef{pull2}. 
Note that  the second fundamental form in the base space satisfy the same  identity as in \reef{Identity}, \ie
 \beqa
\tilde{\Omega }_{\tilde a\tilde b}{}^{\tilde \mu }{\partial _{\tilde c}}{X^{\tilde \nu }}g_{\tilde \mu\tilde \nu} = 0 \,.\labell{Identityred}
\eeqa
However, 
 there is no such relation for $\tilde{\Omega}_{\ta\tb}{}^y$, \ie $\tilde{\Omega}_{\ta\tb}{}^y \prt_\tc S\neq 0$.  Using the above steps, one can calculate ${L}_{p-1}^{(1)t}$. 
 
 If one imposes the constraint that ${L}_{p-1}^{(1)wT}$ is the same as ${L}_{p-1}^{(1)t}$, then one would find the wrong result that all coupling constants in \reef{Lp} are zero. This indicates that the T-duality transformation \reef{a20} must recive higher-derivative corrections.  Since we did not use $X^\mu$-field redefinition to find the independent couplings in \reef{Lp}, the T-duality also has no correction to the last line in \reef{a20}. For the other fields, we consider the following corrections:
\beqa
{A_y} &\rightarrow & S+\sqrt{\alpha'}\delta S,\nonumber\\
{A_{\tilde a}} &\rightarrow & {A_{\tilde a}}+\sqrt{\alpha'}\delta A_\ta\,.\labell{a21}
%{X ^{\tilde \mu }}&\rightarrow &{X ^{\tilde \mu }}\,.
\eeqa
There are the following 16 structures for constructing $\delta A_\ta$:
\beqa
&&\tilde{\Omega }{}^y\Big[F(\prt S+\prt S^3+\prt S^5)+F^3(\prt S+\prt S ^3)+F^5\prt S\Big]\nn\\
&&+\tilde{\nabla}F\Big[1+\prt S^2+\prt S^4+\prt S^6+F^2(1+\prt S^2+\prt S^4)+F^4(1+\prt S^2)+F^6\Big]\,.
\eeqa 
If one considers all contractions of these structures with arbitrary parameters, one would find 253 terms in $\delta A_\ta$. There are the following 16 structures for constructing $\delta S$:
\beqa
&&\tilde{\Omega }{}^y\Big[1+\prt S^2+\prt S^4+\prt S^6+F^2(1+\prt S^2+\prt S ^4)+F^4(1+\prt S^2)+F^6\Big]\nn\\
&&+\tilde{\nabla}F\Big[F(\prt S+\prt S^3+\prt S^5+F^3(\prt S+\prt S^3)+F^5\prt S\Big]\,.
\eeqa 
If one considers all contractions of these structures with arbitrary parameters, one would find 105 terms in $\delta S$. 
 
On the other hand, the transformation of ${L}_p^{(0)w}$ under the above T-duality transformation produces  ${L}_{p-1}^{(0)wT}={L}_{p-1}^{(0)t}$ on the second line of \reef{DBIw}  and the following  couplings at order $\alpha'$:
\beqa
\delta {L}_{p-1}^{(0)wT}=-\frac{T_{p-1}\alpha'}{2}\sqrt{-\det(\tg_{\ta\tb}+F_{\ta\tb}+\prt_\ta S\prt_\tb S)}\Tr\Big[(\tg+F+\prt S \prt S)^{(-1)}(\delta F +\delta (\prt S \prt S)\Big]\,,
\eeqa
where $\delta F_{\ta\tb} = \tilde{\nabla}_\ta \delta A_\tb - \tilde{\nabla}_\tb \delta A_\ta$ and $\delta(\prt_\ta S\prt_\tb S)=\tilde\nabla_\ta\delta S \prt_\tb S+\prt_\ta S \tilde{\nabla}_\tb \delta S$.
One should expand the term inside the bracket, i.e. $\Tr\Big[(\tg+F+\prt S \prt S)^{(-1)}(\delta F +\delta (\prt S \prt S)\Big]$ and consider only the terms up to 8th order of $F$, $\prt S$ and $\tilde{\Omega}^y$.
The above  non-zero term should be added to the T-duality transformation of the Lagrangian at $\mathcal{O}(\alpha')$.

The T-duality constraint at order $\alpha'$ then is the following:
\beqa
{L}_{p-1}^{(1)wT}-{L}_{p-1}^{(1)t}+\delta {L}_{p-1}^{(0)wT}&=&0\,.\labell{LLL}
\eeqa
Since all the above three terms have the DBI factor, we can rewrite the above constraint as
\beqa
\sqrt{-\det(\tg_{\ta\tb}+F_{\ta\tb}+\prt_\ta S\prt_\tb S)}\Delta L&=&0\,.\nn
\eeqa
The overall DBI factor is non-zero, hence the T-duality constraint is
\beqa
\Delta L&=&0\,.\labell{L}
\eeqa
After imposing the Bianchi identities, we solve the above equation to find the coupling constants in \reef{Lp} and the parameters of the T-duality corrections in $\delta A_\ta$ and $\delta S$.

To impose the Bianchi identities corresponding to the field $X^\tmu$, we work in the local inertial frame where the Christoffel connection made from the base space pullback is zero, but its derivative is not. Since there is no covariant derivative of $\tilde{\Omega}_{\ta\tb}{}^\tmu$ in our calculations at $\mathcal{O}(\alpha')$, we write it as $\tilde{\Omega}_{\ta\tb}{}^\tmu=\partial_\ta\partial_\tc X^\tmu$. However, there is $\tilde{\nabla}_\ta \tilde{\Omega}_{\tb\tc}{}^y$, so we write $\tilde{\Omega}_{\tb\tc}{}^y=\tilde{\nabla}_\tb\partial_\tc S$, and then express all covariant derivatives in terms of partial derivatives and Christoffel connections.
The first partial derivative of the Christoffel connection is:
\begin{equation}
\partial_\ta\tilde{\Gamma}^\tb{}_{\tc\td} = \partial_\ta\partial^\tb X^\tmu\partial_\td\partial_\tc X_\tmu + \partial_\ta\partial_\tc\partial_\td X^\tmu\partial^\tb X_\tmu\,,
\end{equation}
where we have used the fact that in the local frame, the first partial derivative of the pullback metric is zero, i.e. $\partial_\ta\tg_{\tb\tc}=0$.
To impose the Bianchi identities corresponding to the $A_a$-field, we express $F$ in terms of the potential $A$ wherever its partial derivative appears. This ensures all Bianchi identities are satisfied, and the equation \reef{L} can be written in an independent but non-gauge invariant form. The coefficient of each independent term up to 8th order in $A, S, X$ must vanish, producing algebraic equations involving the above parameters.

We find the algebraic equations fix all 145 coupling constants in \reef{Lp} in terms of three parameters. However, since we did not use $X^\mu$-field redefinitions and did not perform integration by parts in our calculations, these three parameters may not all be physically significant. In fact, the parameters that can be removed from the couplings by field redefinitions and integration by parts are not physical parameters. So one can safely remove those parameters.
In fact, using the same calculations as in the previous subsection but including the $X^\mu$-field redefinitions as well as integration by parts, we find that one of the remaining three parameters can be removed by field redefinitions and integration by parts. The other two parameters, however, cannot be removed by field redefinitions and integration by parts. Therefore, they are the physical parameters.
To fix these two parameters, we compare the couplings we have found with the couplings that are reported in \cite{Karimi:2018vaf} using T-duality and S-matrix methods, which appear in Appendix A.
The two actions must be identical up to field redefinitions and integration by parts. This process allows us to determine the two physical parameters as well. We obtain the following results for the Lagrangian:
\beqa
\bold{L}_p^{(1)}&=&-T_p\sqrt{-\det(\tG_{ab}+F_{ab})}\Big[- \Omega_{ab\mu } \Omega^{ab\mu }  + 2 F_{a}{}^{c} F^{ab} 
\Omega_{b}{}^{d\mu } \Omega_{cd\mu } -  F_{a}{}^{c} F^{ab} 
F_{d}{}^{f} F^{de} \Omega_{be}{}^{\mu } \Omega_{cf\mu } \nn\\&&- 2 F_{a}{}^{c} F^{ab} F_{b}{}^{d} F_{c}{}^{e} 
\Omega_{d}{}^{f\mu } \Omega_{ef\mu } + \frac{1}{2} F_{a}{}^{c} 
F^{ab} F_{b}{}^{d} F_{c}{}^{e} F_{f}{}^{i} F^{fh} 
\Omega_{dh}{}^{\mu } \Omega_{ei\mu } \nn\\&&+ 2 F_{a}{}^{c} F^{ab} 
F_{b}{}^{d} F_{c}{}^{e} F_{d}{}^{f} F_{e}{}^{h} \Omega_{f}{}^{i\mu 
} \Omega_{hi\mu } + 2 F^{ab} F^{cd} 
\Omega_{ac}{}^{\mu } \Omega_{bd\mu }-  
\frac{10}{3} F_{a}{}^{c} F^{ab} F_{b}{}^{d} F^{ef} \Omega_{ce}{}^{
\mu } \Omega_{df\mu }  \nn\\&&+ \frac{1}{3} F_{a}{}^{c} F^{ab} 
F_{b}{}^{d} F_{e}{}^{h} F^{ef} F_{f}{}^{i} \Omega_{ch}{}^{\mu } 
\Omega_{di\mu }+ \frac{27}{10} 
F_{a}{}^{c} F^{ab} F_{b}{}^{d} F_{c}{}^{e} F_{d}{}^{f} F^{hi} 
\Omega_{eh}{}^{\mu } \Omega_{fi\mu } \nn\\&&+ \frac{3}{2} F_{a}{}^{c} 
F^{ab} F_{b}{}^{d} F_{c}{}^{e} F_{f}{}^{i} F^{fh} 
\Omega_{de}{}^{\mu } \Omega_{hi\mu } -  \nabla_{b}F_{ac} \nabla^{c}F^{ab} + 
\frac{2}{3} F^{ab} F^{cd} \nabla_{c}F_{a}{}^{e} \nabla_{d}F_{be} \nn\\&&
- 2 F_{a}{}^{c} F^{ab} \nabla_{c}F_{de} \nabla^{e}F_{b}{}^{d} + 
F_{a}{}^{c} F^{ab} \nabla_{d}F_{ce} \nabla^{e}F_{b}{}^{d} + 
\frac{1}{30} F_{a}{}^{c} F^{ab} F_{b}{}^{d} F^{ef} 
\nabla_{e}F_{c}{}^{h} \nabla_{f}F_{dh}\nn\\&& -  F_{a}{}^{c} F^{ab} 
F_{d}{}^{f} F^{de} \nabla_{c}F_{b}{}^{h} \nabla_{f}F_{eh} + 
\frac{1}{10} F_{a}{}^{c} F^{ab} F^{de} F^{fh} \nabla_{f}F_{bd} 
\nabla_{h}F_{ce} \nn\\&&+ \frac{1}{2} F_{a}{}^{c} F^{ab} F^{de} F^{fh} 
\nabla_{e}F_{bd} \nabla_{h}F_{cf} -  \frac{1}{2} F_{a}{}^{c} 
F^{ab} F^{de} F^{fh} \nabla_{c}F_{bd} \nabla_{h}F_{ef}\nn\\&& -  
F_{a}{}^{c} F^{ab} F_{d}{}^{f} F^{de} \nabla_{f}F_{ch} 
\nabla^{h}F_{be} + \frac{3}{4} F_{a}{}^{c} F^{ab} F_{b}{}^{d} 
F^{ef} \nabla_{f}F_{eh} \nabla^{h}F_{cd} \nn\\&&+ \frac{5}{3} 
F_{a}{}^{c} F^{ab} F_{b}{}^{d} F^{ef} \nabla_{d}F_{fh} 
\nabla^{h}F_{ce} -  \frac{4}{3} F_{a}{}^{c} F^{ab} F_{b}{}^{d} 
F^{ef} \nabla_{f}F_{dh} \nabla^{h}F_{ce}\nn\\&& + \frac{3}{2} 
F_{a}{}^{c} F^{ab} F_{b}{}^{d} F_{c}{}^{e} \nabla_{e}F_{fh} 
\nabla^{h}F_{d}{}^{f} -  F_{a}{}^{c} F^{ab} F_{b}{}^{d} F_{c}{}^{e} 
\nabla_{f}F_{eh} \nabla^{h}F_{d}{}^{f} \nn\\&&-  \frac{8}{105} 
F_{a}{}^{c} F^{ab} F_{b}{}^{d} F_{c}{}^{e} F_{d}{}^{f} F^{hi} 
\nabla_{h}F_{e}{}^{j} \nabla_{i}F_{fj} + \frac{7}{40} F_{a}{}^{c} 
F^{ab} F_{b}{}^{d} F_{cd} F^{ef} F^{hi} \nabla_{h}F_{e}{}^{j} 
\nabla_{i}F_{fj} \nn\\&&+ \frac{3}{2} F_{a}{}^{c} F^{ab} F_{b}{}^{d} 
F_{c}{}^{e} F_{f}{}^{i} F^{fh} \nabla_{e}F_{d}{}^{j} 
\nabla_{i}F_{hj} -  \frac{1}{10} F_{a}{}^{c} F^{ab} F_{d}{}^{f} 
F^{de} F_{h}{}^{j} F^{hi} \nabla_{i}F_{be} \nabla_{j}F_{cf}\nn\\&& -  
\frac{8}{35} F_{a}{}^{c} F^{ab} F_{b}{}^{d} F_{e}{}^{h} F^{ef} 
F^{ij} \nabla_{i}F_{cf} \nabla_{j}F_{dh} -  \frac{3}{10} 
F_{a}{}^{c} F^{ab} F_{b}{}^{d} F_{e}{}^{h} F^{ef} F^{ij} 
\nabla_{h}F_{cf} \nabla_{j}F_{di} \nn\\&&-  \frac{4}{35} F_{a}{}^{c} 
F^{ab} F_{b}{}^{d} F_{c}{}^{e} F^{fh} F^{ij} \nabla_{i}F_{df} 
\nabla_{j}F_{eh} + \frac{1}{10} F_{a}{}^{c} F^{ab} F_{b}{}^{d} 
F_{c}{}^{e} F^{fh} F^{ij} \nabla_{h}F_{df} \nabla_{j}F_{ei} \nn\\&&-  
\frac{1}{5} F_{a}{}^{c} F^{ab} F_{d}{}^{f} F^{de} F_{h}{}^{j} 
F^{hi} \nabla_{c}F_{be} \nabla_{j}F_{fi} -  \frac{1}{10} 
F_{a}{}^{c} F^{ab} F_{b}{}^{d} F_{c}{}^{e} F^{fh} F^{ij} 
\nabla_{e}F_{df} \nabla_{j}F_{hi}\nn\\&& -  \frac{3}{5} F_{a}{}^{c} 
F^{ab} F_{b}{}^{d} F_{e}{}^{h} F^{ef} F^{ij} \nabla_{f}F_{cd} 
\nabla_{j}F_{hi} -  \frac{2}{5} F_{a}{}^{c} F^{ab} F_{b}{}^{d} 
F_{e}{}^{h} F^{ef} F_{f}{}^{i} \nabla_{i}F_{hj} \nabla^{j}F_{cd} \nn\\&&+ 
\frac{8}{5} F_{a}{}^{c} F^{ab} F_{b}{}^{d} F_{e}{}^{h} F^{ef} 
F_{f}{}^{i} \nabla_{i}F_{dj} \nabla^{j}F_{ch} -  \frac{6}{5} 
F_{a}{}^{c} F^{ab} F_{b}{}^{d} F_{c}{}^{e} F_{f}{}^{i} F^{fh} 
\nabla_{e}F_{ij} \nabla^{j}F_{dh}\nn\\&& + \frac{4}{5} F_{a}{}^{c} 
F^{ab} F_{b}{}^{d} F_{c}{}^{e} F_{f}{}^{i} F^{fh} \nabla_{i}F_{ej} 
\nabla^{j}F_{dh} -  \frac{4}{5} F_{a}{}^{c} F^{ab} F_{b}{}^{d} 
F_{c}{}^{e} F_{d}{}^{f} F^{hi} \nabla_{i}F_{hj} \nabla^{j}F_{ef}\nn\\&& + 
\frac{7}{80} F_{a}{}^{c} F^{ab} F_{b}{}^{d} F_{cd} F^{ef} F^{hi} 
\nabla_{i}F_{hj} \nabla^{j}F_{ef} -  \frac{28}{15} F_{a}{}^{c} 
F^{ab} F_{b}{}^{d} F_{c}{}^{e} F_{d}{}^{f} F^{hi} \nabla_{f}F_{ij} 
\nabla^{j}F_{eh}\nn\\&& + \frac{4}{3} F_{a}{}^{c} F^{ab} F_{b}{}^{d} 
F_{c}{}^{e} F_{d}{}^{f} F^{hi} \nabla_{i}F_{fj} \nabla^{j}F_{eh} -  
\frac{7}{20} F_{a}{}^{c} F^{ab} F_{b}{}^{d} F_{cd} F^{ef} F^{hi} 
\nabla_{i}F_{fj} \nabla^{j}F_{eh}\nn\\&& -  \frac{8}{5} F_{a}{}^{c} 
F^{ab} F_{b}{}^{d} F_{c}{}^{e} F_{d}{}^{f} F_{e}{}^{h} 
\nabla_{h}F_{ij} \nabla^{j}F_{f}{}^{i} -  \frac{7}{40} 
F_{a}{}^{c} F^{ab} F_{b}{}^{d} F_{cd} F_{e}{}^{h} F^{ef} 
\nabla_{h}F_{ij} \nabla^{j}F_{f}{}^{i}\nn\\&& + F_{a}{}^{c} F^{ab} 
F_{b}{}^{d} F_{c}{}^{e} F_{d}{}^{f} F_{e}{}^{h} \nabla_{i}F_{hj} 
\nabla^{j}F_{f}{}^{i}\Big]\,.\labell{Lpf}
\eeqa
The above Lagrangian is invariant under the T-duality \reef{a21}. The corresponding corrections $\delta A_\ta$ and $\delta S$ appear in Appendix B.

It is important to note that all independent terms in \reef{Lp} where the $F$-fields contract with themselves, like the term with coefficient $a_{145}$, are zero. All such couplings are produced solely by expanding the DBI factor in the above Lagrangian.
Interestingly, if one replaces the DBI factor in \reef{Lp} with $\sqrt{-\det(\tG)}$, the basis remains covariant and contains the same 145 terms. However, the corresponding fixed couplings after performing T-duality and comparing with the effective action \cite{Karimi:2018vaf} would result in 104 couplings, instead of the 49 couplings in \reef{Lpf}. This suggests that the choice of the DBI factor can have a significant impact on the structure and number of the resulting couplings.
We will discuss this point in more detail in the discussion section.

\section{Boundary Lagrangian}

When a $D_p$-brane is entirely embedded in the boundary, the world-volume theory should include all the massless fields as well as the unit vector field orthogonal to the boundary. The pull-back of the massless closed string fields as well as the first derivative of the massless open string fields appear nonlinearly in the DBI action. This raises the question of whether it is possible to include the normal vector and its first derivative, which is the extrinsic curvature, into the DBI action. To find the answer to this question, one may consider T-duality on massless closed string fields, or T-duality  on  massless open string fields.

This question has been addressed in \cite{Hosseini:2022vrr} by considering the massless closed string fields. It has been argued in \cite{Hosseini:2022vrr} that T-duality does not allow terms such as $n \tilde{B} \nabla \tilde{B}$, $n \tilde{B} \tilde{B} \tilde{B} \nabla \tilde{B}$, and so on, in the boundary action. The terms that include the trace of the pull-back of the extrinsic curvature seem to arise from incorporating the extrinsic curvature under the square root in the DBI action and subsequently expanding it. Additionally, due to a restriction on selecting the base space background such that the circular reduction should have the symmetry $U(1) \times U(1)$ \cite{Hosseini:2022vrr}, the couplings in which the indices of the extrinsic curvature are contracted with $\tilde{B}$'s cannot be fixed by T-duality. By examining the T-duality of massless open  string fields in this section, we find that such couplings can also be derived by expanding the following extension of the DBI action:
\beqa
\prt\bold{L}_p&=&-T_p\, e^{-\Phi}\sqrt{-\det(\tG_{ab}+\tB_{ab}+F_{ab}+2\sqrt{\alpha'}\prt_a X^\mu\prt_b X^\nu K_{\mu\nu})}\,.\labell{DBIK1}
\eeqa
where $K_{\mu\nu}$ is given in \reef{ext}, in which the covariant derivative is replaced by the partial derivative. We assume the closed string fields are constant, and the partial derivative of the normal vector to the boundary is not zero, i.e., $K_{\mu\nu} \neq 0$.

To demonstrate that the above boundary Lagrangian is invariant under T-duality, we consider the following circular reduction for the closed string fields:
\beqa
G_{\mu\nu}=
\left(\matrix{
\bg_{\tmu \tnu} & 0\!\!\!\!\! &\cr
0 &e^{\vp} \!\!\!\!\!&
}\right)\,,\,B_{\mu\nu}=
\left(\matrix{
\bb_{\tmu \tnu} & 0\!\!\!\!\! &\cr
0 &0 \!\!\!\!\!&
}\right)\,,\,\Phi=\bphi+\vp/4\,,
\eeqa
where $\bg_{\tmu \tnu}$, $\bb_{\tmu \tnu}$, and $\bphi$ are the base space fields that are invariant under T-duality. The base space scalar field $\vp$ transforms under T-duality as
\beqa
\vp\rightarrow -\vp\,.
\eeqa
Since there is no vector field in these reductions, the corresponding reduction of the Lagrangian  in \reef{DBIK1} has the symmetry $U(1) \times U(1)$. The above reduction produces the following reduction for $K_{\mu\nu}$:
\beqa
K_{\mu\nu}=
\left(\matrix{
\bK_{\tmu \tnu} & 0\!\!\!\!\! &\cr
0 &0 \!\!\!\!\!&
}\right),
\eeqa
where $\bK_{\tmu\tnu}=\prt_\tmu n_{\tnu}-n_\tmu n^{\tilde{\rho}}\prt_{\tilde{\rho}}n_{\tnu}$ is the extrinsic curvature in the base space, which  is invariant under T-duality.

The world-volume reduction of the $D_p$-brane Lagrangian in \reef{DBIK1} then becomes
\beqa
\prt{L}_p^w=-T_p\, e^{-\bphi+\vp/4}\sqrt{-\det\Big[\prt_\ta X^\tmu\prt_\tb X^\tnu(\bg_{\tmu\tnu}+\bb_{\tmu\tnu}+2\sqrt{\alpha'}\bK_{\tmu\tnu})+F_{\ta\tb}+e^{-\vp}\prt_\ta A_y\prt_\tb A_y\Big]}\,.\labell{DBIK2}
\eeqa
And the transverse reduction of the $D_{p-1}$-brane Lagrangian  in \reef{DBIK1} becomes
\beqa
\prt{L}_{p-1}^t=-T_{p-1}\, e^{-\bphi-\vp/4}\sqrt{-\det\Big[\prt_\ta X^\tmu\prt_\tb X^\tnu(\bg_{\tmu\tnu}+\bb_{\tmu\tnu}+2\sqrt{\alpha'}\bK_{\tmu\tnu})+F_{\ta\tb}+e^{\vp}\prt_\ta S\prt_\tb S\Big]}\,.\labell{DBIK3}
\eeqa
The transformation of $\prt{L}_p^w$ under T-duality \reef{a20} becomes identical to $\prt{L}_{p-1}^t$.

The above T-duality symmetry does not fix the coefficient of $K_{\mu\nu}$ in the modified DBI action in \reef{DBIK1}. This coefficient is fixed in \cite{Hosseini:2022vrr} by imposing T-duality on the combination of bulk and boundary actions when the $D_p$-brane extends in the bulk of spacetime and ends on the boundary. In this case, the bulk of the D-brane is T-dual up to some total derivative term in the base space. These anomalous total derivative terms then transfer to the boundary by using the Stokes theorem. Then, the boundary couplings should be such that their T-duality transformation cancels the anomalous term. In this way, the boundary coupling produces the extrinsic curvature that is in harmony with the bulk action. The coefficient of the extrinsic curvature in \reef{DBIK1} is the one fixed in \cite{Hosseini:2022vrr}.

\section{Discussion}

When a $D_p$-brane is in a spacetime with boundary and the $D_p$-brane extends in the bulk and ends on the boundary, the world-volume Lagrangian has two parts: the bulk Lagrangian and the boundary Lagrangian. Each of these Lagrangians has an $\alpha'$ expansion that may be found by imposing the T-duality constraint on the Lagrangians.  The T-duality of the world-volume reduction of each $D_p$-brane Lagrangian must be the same as the transverse reduction of the $D_{p-1}$-brane Lagrangian. We have done this calculation for the bulk Lagrangian at order $\alpha'$ and found an expansion for the couplings up to 8th order of the Maxwell field strength $F$ and the second fundamental form $\Omega$. This Lagrangian is given in \reef{Lpf}  and is invariant under T-duality, as shown in \reef{a21} with the corrections that appear in Appendix B.

To find the boundary Lagrangian, we considered the case where the $D_p$-brane is entirely along the boundary of spacetime and showed that the modified DBI action in \reef{DBIK} is invariant under the standard leading-order T-duality transformation. The higher derivative corrections to this Lagrangian can be divided into two parts. One part involves exactly the same higher derivative corrections as in the DBI action, some of which have been found in \reef{Lpf}. The other part includes couplings in which the vector $n_\mu$ or its derivatives appear in each coupling. In other words, when a $D_p$-brane extends in the bulk and ends on the boundary, the effective actions should be
\beqa
\bS_p^D+\prt\!\!\bS_p^D&\!\!\!\!\!\!=\!\!\!\!\!\!&-T_p \int d^{p+1}\sigma\, e^{-\Phi}\sqrt{-\det(\tG_{ab}+\tB_{ab}+F_{ab})}[1+O_{bulk}(\alpha')]\\
&&-T_p\sqrt{\alpha'}\int d^{p}\tau\,e^{-\Phi}\sqrt{-\det(\tG_{\ba\bb}\!+\!\tB_{\ba\bb}\!+\!F_{\ba\bb}\!+\!2\sqrt{\alpha'}\tilde{K}_{\ba\bb})}[1\!+\!O_{bulk}(\alpha')\!+\!O_n(\alpha')].\nn
\eeqa
 The corrections $O_{bulk}(\alpha')$ for the bulk and boundary are the same. In the pull-back of the spacetime fields in the bulk, $X^\mu=X^\mu(\sigma)$, and in the boundary, $X^\mu=X^\mu(\tau)$. The bulk world-volume coordinates are $\sigma^0,\cdots ,\sigma^p$, and the boundary world-volume coordinates are $\tau^0,\cdots,\tau^{p-1}$. Each term in the $\alpha'$-corrections $O_n(\alpha')$ should involve the unit vector $n_{\mu}$.

For the $O_p$-plane, there is no $\tilde{B}$ and no open string fields. So the $O_p$-plane effective action  is:
\beqa
\bS_p^O+\prt\!\!\bS_p^O&\!\!\!\!\!=\!\!\!\!\!&-T_p \int d^{p+1}\sigma\, e^{-\Phi}\sqrt{-\det(\tG_{ab})}[1+O_{bulk}(\alpha')]\\
&&-T_p\sqrt{\alpha'}\int d^{p}\tau\,e^{-\Phi}\sqrt{-\det(\tG_{\ba\bb}+2\sqrt{\alpha'}\tilde{K}'_{\ba\bb})}[1\!+\!O_{bulk}(\alpha')\!+\!O_n(\alpha')].\nn
\eeqa
Using the same steps as in Section 3, one can easily observe that the leading order term in   $O_p$-plane boundary Lagrangian is invariant under T-duality. 

We have observed that the bulk $D_p$-brane Lagrangian at order $\alpha'$ is invariant under T-duality without any residual total derivative terms in the base space, provided that one does not use the $X^\mu$-field redefinition in producing the independent basis. This may be a consequence of the fact that, while the effective action is invariant under coordinate transformations, there is no such symmetry for the effective Lagrangian. That is
\beqa
\int d^{p+1}\sigma\sqrt{-\det(\tG_{ab})}&\rightarrow &\int d^{p+1}\sigma\sqrt{-\det(\tG_{ab})}\,,\nn\\
\sqrt{-\det(\tG_{ab})}&\not\rightarrow &\sqrt{-\det(\tG_{ab})}\,.
\eeqa
This indicates that T-duality of the Lagrangian is satisfied in specific schemes. A similar observation has been recently found in studying the T-duality of the effective action of the heterotic theory, which includes both NS-NS and YM fields. The T-duality is not satisfied for any specific scheme. In fact, it has been observed that T-duality is not satisfied by the minimal basis in which all redundant couplings related by all field redefinitions are removed. However, the T-duality is satisfied by the maximal basis in which redundant terms due to the field redefinitions are not removed \cite{Garousi:2024vbz,Garousi:2024imy}.

The T-duality invariant Lagrangian in  \reef{Lpf} can be expressed in several other T-duality invariant forms by employing $A_a$-field redefinitions. On the other hand, it is known that the pull-back of the $B$-field, denoted as $\tilde{B}$, appears in the $D$-brane Lagrangian by replacing $F \rightarrow F + \tilde{B}$. This suggests that there should be a $\tilde{B}$-field redefinition in the world-volume action in addition to the $B$-field redefinition in the spacetime action. In fact, $\tilde{B}$ satisfies the following Bianchi identity \cite{Hosseini:2022vrr}:
\beqa
\tilde{\nabla}_{[a}\tB_{bc]}&=&0\,.
\eeqa
This indicates that one is free to add $\partial_{[a}\mathcal{A}_{b]}$ to $\tilde{B}$, where $\mathcal{A}_a$ is an arbitrary vector. Hence, $\tilde{B}_{ab}$, like $F_{ab}$, has the following field redefinition at order $\alpha'$:
\beqa
\tB_{ab}\rightarrow \tB_{ab}+\alpha'\prt_{[a}\cA^{(1)}_{b]}\,,
\eeqa 
where $\mathcal{A}^{(1)}_a$ is an arbitrary vector constructed from the massless fields at order $\sqrt{\alpha'}$. Such a field redefinition has no effect on the $B$-field strength $H$ that appears in the spacetime action, as the exterior derivative of the exterior derivative is zero, i.e., $d^2 = 0$.

We have found the T-duality invariant couplings \reef{Lpf} at order $\alpha'$ up to the 8th order of the dimensionless field strength $F$ for the particular basis \reef{Lp} in which the covariant derivatives are constructed from the pull-back metric $\tG_{ab}$ and indices are also contracted with inverse of this metric. In principle, one may extend this $\tG$-covariant basis to include all orders of $F$ and impose T-duality to find the corresponding coupling constants.
This raises the question of whether it is possible to find a particular scheme in which all orders of $F$ can be written in a closed-form expression. One approach to finding such a scheme might involve considering the Lagrangian \reef{Lpf}, which features an overall DBI factor. This factor is given by the square root of the inverse of the following matrix:
\beqa
\left(\frac{1}{\tG+F}\right)^{ab}&=&\cg^{ab}-\Theta ^{ab}\,,
\eeqa
where the symmetric and antisymmetric parts are, respectively,
\beqa
\cg^{ab}=\left(\frac{1}{\tG+F}\tG\frac{1}{\tG-F}\right)^{ab}\,,\nn&&
\Theta ^{ab}=\left(\frac{1}{\tG+F}F\frac{1}{\tG-F}\right)^{ab}\,.
\eeqa
The symmetric part is the inverse of the following metric:
\beqa
\cg_{ab}&=&\tG_{ab}-F_{ac}F^c{}_b\,.
\eeqa
On the other hand, if one chooses another scheme for the basis \reef{Lp}, it is possible to change the coefficient of the last term in the second line to be 2, and remove the first term in the fifth line of \reef{Lpf}. Then all terms in \reef{Lpf} with the structure $\Omega_{ab}{}^\mu\Omega_{cd\mu}$ whose world-volume indices are contracted with an even number of $F$, i.e., all terms in the first and second lines and the first term in the third line, can be written as $-\cg^{ac}\cg^{bd}\Omega_{ab}{}^\mu\Omega_{cd\mu}$.
Therefore, one may expect that the world-volume covariant derivative constructed from the above metric, i.e., $\nabla_{\cg}$, may be the appropriate covariant derivative to consider for constructing the second fundamental form and constructing the independent basis in which all indices are contracted with $\cg^{ab}$ and $\Theta^{ab}$. Then T-duality can fix the coupling constants of such a $\cg$-covariant basis. It would be interesting to perform the appropriate calculation to determine the coupling constants of the $\cg$-covariant basis. It may be possible that the infinite number of couplings at order $\alpha'$ can be expressed as a few couplings constructed from the $\cg$-metric and $\Theta^{ab}$.

{\bf Acknowledgements}:  This work is supported by Ferdowsi University of Mashhad under grant  332(1401/12/10).

%The covariant  result obviously can not be written in terms of metric V that appear in string theory S-matrix. 

%\newpage
\vskip 0.5cm
{\Large \bf Appendix A: Effective Action}
%\vskip 0.5 cm

In this Appendix, we write the effective action of $D_p$-brane at order $\alpha'$ and up to 8th order of $F$ in the closed spacetime manifold that has been found in  \cite{Karimi:2018vaf}. The basis which includes all gauge-invariant couplings modulo the most general field redefinition, integration by parts, and the use of Bianchi identities has 81 couplings. T-duality fixes them up to two parameters, and the two remaining parameters are also fixed by comparing the four-point function of the resulting effective action with the corresponding disk-level S-matrix elements. This  action has   the following 79 couplings \cite{Karimi:2018vaf}:
\beqa
\bS_p^{(1)}&\!\!\!\!\!\!\!\!=\!\!\!\!\!\!\!\!&-T_p\alpha'\int d^{p+1}\sigma\sqrt{-\det(\tG_{ab})}\Big[- \Omega_{ab\mu } \Omega^{ab\mu } + F^{ab} F^{cd} 
\Omega_{ac}{}^{\mu } \Omega_{bd\mu } + \Omega^{a}{}_{a}{}^{\mu 
} \Omega^{b}{}_{b\mu }\nn\\&& + F_{a}{}^{c} F^{ab} \Omega_{b}{}^{d\mu } 
\Omega_{cd\mu } -  \frac{1}{3} F_{a}{}^{c} F^{ab} F_{d}{}^{f} 
F^{de} \Omega_{be}{}^{\mu } \Omega_{cf\mu } -  \frac{1}{4} 
F_{ab} F^{ab} \Omega_{cd\mu } \Omega^{cd\mu }\nn\\&& - \! \frac{4}{3} 
F_{a}{}^{c} F^{ab} F_{b}{}^{d} F^{ef} \Omega_{ce}{}^{\mu } 
\Omega_{df\mu } \!+\! \frac{1}{3} F_{ab} F^{ab} F^{cd} F^{ef} 
\Omega_{ce}{}^{\mu } \Omega_{df\mu } \!+\! \frac{14}{15} 
F_{a}{}^{c} F^{ab} F_{b}{}^{d} F_{e}{}^{h} F^{ef} F_{f}{}^{i} 
\Omega_{ch}{}^{\mu } \Omega_{di\mu } \nn\\&&-  \frac{2}{3} F_{a}{}^{c} 
F^{ab} F_{d}{}^{f} F^{de} \Omega_{bc}{}^{\mu } \Omega_{ef\mu } -  
\frac{2}{3} F_{a}{}^{c} F^{ab} F_{b}{}^{d} F_{c}{}^{e} 
\Omega_{d}{}^{f\mu } \Omega_{ef\mu } + \frac{1}{3} F_{ab} 
F^{ab} F_{c}{}^{e} F^{cd} \Omega_{d}{}^{f\mu } \Omega_{ef\mu } \nn\\&&+ 
\frac{3}{5} F_{a}{}^{c} F^{ab} F_{b}{}^{d} F_{c}{}^{e} F_{f}{}^{i} 
F^{fh} \Omega_{dh}{}^{\mu } \Omega_{ei\mu } \!+\! \frac{1}{8} 
F_{a}{}^{c} F^{ab} F_{b}{}^{d} F_{cd} \Omega_{ef\mu } 
\Omega^{ef\mu } \!-\!  \frac{1}{32} F_{ab} F^{ab} F_{cd} F^{cd} 
\Omega_{ef\mu } \Omega^{ef\mu }\nn\\&& + \frac{6}{5} F_{a}{}^{c} 
F^{ab} F_{b}{}^{d} F_{c}{}^{e} F_{d}{}^{f} F^{hi} 
\Omega_{eh}{}^{\mu } \Omega_{fi\mu } -  \frac{1}{3} F_{ab} 
F^{ab} F_{c}{}^{e} F^{cd} F_{d}{}^{f} F^{hi} \Omega_{eh}{}^{\mu } 
\Omega_{fi\mu }\nn\\&& -  \frac{1}{8} F_{a}{}^{c} F^{ab} F_{b}{}^{d} 
F_{cd} F^{ef} F^{hi} \Omega_{eh}{}^{\mu } \Omega_{fi\mu } + 
\frac{1}{32} F_{ab} F^{ab} F_{cd} F^{cd} F^{ef} F^{hi} 
\Omega_{eh}{}^{\mu } \Omega_{fi\mu } \nn\\&&+ \frac{16}{15} 
F_{a}{}^{c} F^{ab} F_{b}{}^{d} F_{c}{}^{e} F_{f}{}^{i} F^{fh} 
\Omega_{de}{}^{\mu } \Omega_{hi\mu } -  \frac{1}{12} F_{ab} 
F^{ab} F_{c}{}^{e} F^{cd} F_{f}{}^{i} F^{fh} \Omega_{de}{}^{\mu } 
\Omega_{hi\mu } \nn\\&&+ F_{a}{}^{c} F^{ab} F_{b}{}^{d} F_{c}{}^{e} 
F_{d}{}^{f} F_{e}{}^{i} \Omega_{f}{}^{h\mu } \Omega_{hi\mu } -  
\frac{1}{4} F_{ab} F^{ab} F_{c}{}^{e} F^{cd} F_{d}{}^{f} 
F_{e}{}^{i} \Omega_{f}{}^{h\mu } \Omega_{hi\mu }\nn\\&& -  \frac{1}{8} 
F_{a}{}^{c} F^{ab} F_{b}{}^{d} F_{cd} F_{e}{}^{i} F^{ef} 
\Omega_{f}{}^{h\mu } \Omega_{hi\mu } + \frac{1}{32} F_{ab} 
F^{ab} F_{cd} F^{cd} F_{e}{}^{i} F^{ef} \Omega_{f}{}^{h\mu } 
\Omega_{hi\mu } \nn\\&&-  \frac{1}{12} F_{a}{}^{c} F^{ab} F_{b}{}^{d} 
F_{c}{}^{e} F_{d}{}^{f} F_{ef} \Omega_{hi\mu } \Omega^{hi\mu } + 
\frac{1}{32} F_{ab} F^{ab} F_{c}{}^{e} F^{cd} F_{d}{}^{f} F_{ef} 
\Omega_{hi\mu } \Omega^{hi\mu }\nn\\&& -\!  \frac{1}{384} F_{ab} F^{ab} 
F_{cd} F^{cd} F_{ef} F^{ef} \Omega_{hi\mu } \Omega^{hi\mu } \!+ \!
\frac{1}{6} F_{a}{}^{c} F^{ab} \nabla_{b}F^{de} \nabla_{c}F_{de} \!
-  \!\frac{1}{15} F_{a}{}^{c} F^{ab} F_{b}{}^{d} F^{ef} 
\nabla_{c}F_{e}{}^{i} \nabla_{d}F_{fi}\nn\\&& -  \frac{4}{5} F_{a}{}^{c} 
F^{ab} F_{b}{}^{d} F_{e}{}^{i} F^{ef} F^{hj} \nabla_{c}F_{fh} 
\nabla_{d}F_{ij} -  \frac{3}{5} F_{a}{}^{c} F^{ab} F_{d}{}^{f} 
F^{de} \nabla_{c}F_{fi} \nabla_{e}F_{b}{}^{i}\nn\\&& -  \frac{1}{12} 
F_{ab} F^{ab} F^{cd} F^{ef} \nabla_{d}F_{fi} \nabla_{e}F_{c}{}^{i} 
-  \frac{1}{3} F_{a}{}^{c} F^{ab} \nabla_{e}F_{cd} 
\nabla^{e}F_{b}{}^{d} -  \frac{1}{24} F_{ab} F^{ab} 
\nabla_{e}F_{cd} \nabla^{e}F^{cd}\nn\\&& + F_{a}{}^{c} F^{ab} F^{de} 
F^{fi} \nabla_{c}F_{ei} \nabla_{f}F_{bd} + \frac{3}{5} 
F_{a}{}^{c} F^{ab} F^{de} F^{fi} \nabla_{e}F_{ci} \nabla_{f}F_{bd} 
+ \frac{3}{5} F_{a}{}^{c} F^{ab} F_{d}{}^{f} F^{de} 
\nabla_{c}F_{b}{}^{i} \nabla_{f}F_{ei}\nn\\&& -  \frac{6}{5} F_{a}{}^{c} 
F^{ab} F_{b}{}^{d} F_{c}{}^{e} F_{f}{}^{i} F^{fh} 
\nabla_{e}F_{d}{}^{j} \nabla_{h}F_{ij} + \frac{2}{15} F_{ab} 
F^{ab} F_{c}{}^{e} F^{cd} F_{f}{}^{i} F^{fh} \nabla_{e}F_{d}{}^{j} 
\nabla_{h}F_{ij} \nn\\&&+ F_{a}{}^{c} F^{ab} F_{b}{}^{d} F_{e}{}^{h} F^{ef} 
F_{f}{}^{i} \nabla_{h}F_{c}{}^{j} \nabla_{i}F_{dj} + \frac{8}{5} 
F_{a}{}^{c} F^{ab} F^{de} F^{fi} \nabla_{c}F_{bd} \nabla_{i}F_{ef} \nn\\&&
-  \frac{13}{5} F_{a}{}^{c} F^{ab} F_{b}{}^{d} F_{c}{}^{e} F^{fi} 
F^{hj} \nabla_{h}F_{df} \nabla_{i}F_{ej} + \frac{2}{15} F_{ab} 
F^{ab} F_{c}{}^{e} F^{cd} F^{fi} F^{hj} \nabla_{h}F_{df} 
\nabla_{i}F_{ej}\nn\\&& + \frac{6}{5} F_{a}{}^{c} F^{ab} F_{b}{}^{d} 
F_{c}{}^{e} F_{f}{}^{i} F^{fh} \nabla_{h}F_{d}{}^{j} 
\nabla_{i}F_{ej} -  \frac{1}{20} F_{ab} F^{ab} F_{c}{}^{e} F^{cd} 
F_{f}{}^{i} F^{fh} \nabla_{h}F_{d}{}^{j} \nabla_{i}F_{ej}\nn\\&& + 
\frac{7}{10} F_{a}{}^{c} F^{ab} F_{b}{}^{d} F_{e}{}^{i} F^{ef} 
F^{hj} \nabla_{h}F_{cd} \nabla_{i}F_{fj} -  \frac{71}{105} 
F_{a}{}^{c} F^{ab} F_{b}{}^{d} F_{c}{}^{e} F_{d}{}^{f} F^{hi} 
\nabla_{h}F_{e}{}^{j} \nabla_{i}F_{fj}\nn\\&& -  \frac{7}{120} 
F_{a}{}^{c} F^{ab} F_{b}{}^{d} F_{cd} F^{ef} F^{hi} 
\nabla_{h}F_{e}{}^{j} \nabla_{i}F_{fj} -  \frac{7}{480} F_{ab} 
F^{ab} F_{cd} F^{cd} F^{ef} F^{hi} \nabla_{h}F_{e}{}^{j} 
\nabla_{i}F_{fj}\nn\\&& -  \frac{3}{5} F_{a}{}^{c} F^{ab} F_{d}{}^{f} 
F^{de} \nabla_{f}F_{ci} \nabla^{i}F_{be} \!- \! \frac{4}{5} 
F_{a}{}^{c} F^{ab} F_{b}{}^{d} F^{ef} \nabla_{f}F_{di} 
\nabla^{i}F_{ce} + \frac{1}{6} F_{ab} F^{ab} F^{cd} F^{ef} 
\nabla_{f}F_{di} \nabla^{i}F_{ce} \nn\\&&-  \frac{3}{10} F_{a}{}^{c} 
F^{ab} F_{b}{}^{d} F^{ef} \nabla_{d}F_{ci} \nabla^{i}F_{ef} + 
\frac{27}{35} F_{a}{}^{c} F^{ab} F_{b}{}^{d} F_{e}{}^{i} F^{ef} 
F^{hj} \nabla_{h}F_{cf} \nabla_{j}F_{di} \nn\\&&-  \frac{8}{15} 
F_{a}{}^{c} F^{ab} F_{b}{}^{d} F_{c}{}^{e} F^{fi} F^{hj} 
\nabla_{i}F_{df} \nabla_{j}F_{eh} -  \frac{7}{60} F_{ab} F^{ab} 
F_{c}{}^{e} F^{cd} F^{fi} F^{hj} \nabla_{i}F_{df} \nabla_{j}F_{eh} 
\nn\\&&+ \frac{59}{35} F_{a}{}^{c} F^{ab} F_{b}{}^{d} F_{c}{}^{e} F^{fi} 
F^{hj} \nabla_{h}F_{df} \nabla_{j}F_{ei} -  \frac{1}{15} F_{ab} 
F^{ab} F_{c}{}^{e} F^{cd} F^{fi} F^{hj} \nabla_{h}F_{df} 
\nabla_{j}F_{ei}\nn\\&& -  \frac{12}{5} F_{a}{}^{c} F^{ab} F_{b}{}^{d} 
F_{c}{}^{e} F_{f}{}^{i} F^{fh} \nabla_{h}F_{d}{}^{j} 
\nabla_{j}F_{ei} + \frac{14}{15} F_{a}{}^{c} F^{ab} F_{d}{}^{f} 
F^{de} F^{hi} F_{i}{}^{j} \nabla_{c}F_{be} \nabla_{j}F_{fh}\nn\\&& + 
\frac{16}{15} F_{a}{}^{c} F^{ab} F_{b}{}^{d} F_{c}{}^{e} 
F_{d}{}^{f} F^{hi} \nabla_{h}F_{e}{}^{j} \nabla_{j}F_{fi} + 
\frac{31}{15} F_{a}{}^{c} F^{ab} F_{b}{}^{d} F_{c}{}^{e} F^{fi} 
F^{hj} \nabla_{e}F_{df} \nabla_{j}F_{hi}\nn\\&& -  \frac{1}{8} F_{ab} 
F^{ab} F_{c}{}^{e} F^{cd} F_{d}{}^{f} F^{hi} \nabla_{f}F_{e}{}^{j} 
\nabla_{j}F_{hi} -  \frac{13}{10} F_{a}{}^{c} F^{ab} F_{b}{}^{d} 
F_{e}{}^{h} F^{ef} F_{f}{}^{i} \nabla_{j}F_{di} \nabla^{j}F_{ch} \nn\\&&+ 
\frac{7}{120} F_{ab} F^{ab} F_{c}{}^{e} F^{cd} F_{f}{}^{i} F^{fh} 
\nabla_{j}F_{ei} \nabla^{j}F_{dh} -  \frac{8}{15} F_{a}{}^{c} 
F^{ab} F_{b}{}^{d} F_{c}{}^{e} F_{d}{}^{f} F^{hi} \nabla_{j}F_{fi} 
\nabla^{j}F_{eh} \nn\\&&-  \frac{3}{80} F_{a}{}^{c} F^{ab} F_{b}{}^{d} 
F_{cd} F^{ef} F^{hi} \nabla_{j}F_{fi} \nabla^{j}F_{eh} + 
\frac{1}{5} F_{a}{}^{c} F^{ab} F_{b}{}^{d} F_{c}{}^{e} F_{d}{}^{f} 
F_{e}{}^{i} \nabla_{h}F_{ij} \nabla^{j}F_{f}{}^{h}\nn\\&& + \frac{1}{12} 
F_{ab} F^{ab} F_{c}{}^{e} F^{cd} F_{d}{}^{f} F_{e}{}^{i} 
\nabla_{h}F_{ij} \nabla^{j}F_{f}{}^{h} + \frac{1}{10} F_{a}{}^{c} 
F^{ab} F_{b}{}^{d} F_{cd} F_{e}{}^{i} F^{ef} \nabla_{h}F_{ij} 
\nabla^{j}F_{f}{}^{h} \nn\\&&+ \frac{1}{240} F_{ab} F^{ab} F_{cd} F^{cd} 
F_{e}{}^{i} F^{ef} \nabla_{h}F_{ij} \nabla^{j}F_{f}{}^{h} + 
\frac{7}{5} F_{a}{}^{c} F^{ab} F_{b}{}^{d} F_{c}{}^{e} F_{d}{}^{f} 
F_{e}{}^{i} \nabla_{j}F_{hi} \nabla^{j}F_{f}{}^{h}\nn\\&& -  
\frac{1}{30} F_{ab} F^{ab} F_{c}{}^{e} F^{cd} F_{d}{}^{f} 
F_{e}{}^{i} \nabla_{j}F_{hi} \nabla^{j}F_{f}{}^{h} + \frac{1}{4} 
F_{a}{}^{c} F^{ab} F_{b}{}^{d} F_{cd} F_{e}{}^{i} F^{ef} 
\nabla_{j}F_{hi} \nabla^{j}F_{f}{}^{h}\nn\\&& + \frac{7}{160} F_{ab} 
F^{ab} F_{cd} F^{cd} F_{e}{}^{i} F^{ef} \nabla_{j}F_{hi} 
\nabla^{j}F_{f}{}^{h} + \frac{1}{45} F_{a}{}^{c} F^{ab} 
F_{b}{}^{d} F_{c}{}^{e} F_{d}{}^{f} F_{ef} \nabla_{h}F_{ij} 
\nabla^{j}F^{hi}\nn\\&& -  \frac{1}{40} F_{ab} F^{ab} F_{c}{}^{e} F^{cd} 
F_{d}{}^{f} F_{ef} \nabla_{h}F_{ij} \nabla^{j}F^{hi} -  
\frac{1}{2880} F_{ab} F^{ab} F_{cd} F^{cd} F_{ef} F^{ef} 
\nabla_{h}F_{ij} \nabla^{j}F^{hi}\Big]\,.
\eeqa
The above action is invariant under T-duality transformations that have some corrections at order $\alpha'$ for $\delta A_\ta$, $\delta S$, and $\delta X^{\mu}$. There are also some anomalous total derivative terms in the base space that are not invariant under T-duality. However, for a closed spacetime manifold that has no boundary, these anomalous total derivative terms become zero.

Note that in the above scheme, the leading-order propagators of massless open string fields do not receive $\alpha'$ corrections. This makes the comparison of the four-point function of the above effective action with the corresponding disk-level S-matrix element straightforward.
However, the basis that we have chosen in \reef{Lp} corrects the leading-order propagators. So instead of comparing our result with the S-matrix elements, we compare the effective Lagrangian that we have found in this paper with the above effective action. The two must be the same up to $X^\mu$-, $A_a$-field redefinitions, integration by parts, and Bianchi identities.

\vskip 1 cm
{\Large \bf Appendix B: T-Duality Transformations at $\mathcal{O}(\alpha')$}
\vskip 0.5 cm

The T-duality constraint \reef{L} fixes the 145 parameters in \reef{Lp} up to three parameters. The corresponding T-duality transformations are also fixed in terms of these three parameters. The T-duality transformations corresponding to the Lagrangian \reef{Lpf} are as follows:
\beqa
\delta S&\!\!\!\!\!\!\!=\!\!\!\!\!\!\!&\frac{1}{2} dS^{a} dS^{b} F_{a}{}^{c} F_{c}{}^{d} \tilde{\Omega}_{bd} + 
\frac{3}{10} dS^{a} dS^{b} F_{a}{}^{c} F_{c}{}^{d} F_{d}{}^{e} 
F_{e}{}^{f} \tilde{\Omega}_{bf} -  \frac{1}{2} dS^{a} dS^{b} F_{a}{}^{c} 
F_{b}{}^{d} \tilde{\Omega}_{cd}\nn\\&& + \frac{1}{2} dS^{a} dS^{b} dS^{c} 
dS^{d} F_{a}{}^{e} F_{be} \tilde{\Omega}_{cd} + dS_{a} dS^{a} F_{b}{}^{d} 
F^{bc} \tilde{\Omega}_{cd} + \frac{7}{10} dS_{a} dS^{a} dS^{b} dS^{c} 
F_{b}{}^{d} F_{d}{}^{e} \tilde{\Omega}_{ce}\nn\\&& -  \frac{1}{5} dS_{a} dS^{a} 
dS^{b} dS^{c} F_{b}{}^{d} F_{c}{}^{e} \tilde{\Omega}_{de} + \frac{1}{2} 
F_{a}{}^{c} F^{ab} F_{b}{}^{d} F_{c}{}^{e} \tilde{\Omega}_{de} + 
\frac{3}{20} dS_{a} dS^{a} dS_{b} dS^{b} F_{c}{}^{e} F^{cd} \tilde{\Omega}_{de} \nn\\&&+ \frac{3}{10} dS^{a} dS^{b} F_{a}{}^{c} F_{b}{}^{d} 
F_{c}{}^{e} F_{e}{}^{f} \tilde{\Omega}_{df} -  \frac{3}{5} dS^{a} dS^{b} 
F_{a}{}^{c} F_{b}{}^{d} F_{c}{}^{e} F_{d}{}^{f} \tilde{\Omega}_{ef} -  
\frac{3}{10} dS_{a} dS^{a} F_{b}{}^{d} F^{bc} F_{c}{}^{e} 
F_{d}{}^{f} \tilde{\Omega}_{ef}\nn\\&& + \frac{1}{2} dS^{a} dS^{b} F_{a}{}^{c} 
F_{bc} F_{d}{}^{f} F^{de} \tilde{\Omega}_{ef} -  \frac{2}{5} F_{a}{}^{c} 
F^{ab} F_{b}{}^{d} F_{c}{}^{e} F_{d}{}^{f} F_{e}{}^{h} \tilde{\Omega}_{fh} 
+ \frac{7}{40} F_{a}{}^{c} F^{ab} F_{b}{}^{d} F_{cd} F_{e}{}^{h} 
F^{ef} \tilde{\Omega}_{fh} \nn\\&&-  dS^{a} dS^{b} dS^{c} F_{a}{}^{d} F_{d}{}^{e} 
F_{e}{}^{f} \nabla_{c}F_{bf} + \frac{1}{10} dS^{a} F_{a}{}^{b} 
F_{b}{}^{c} F_{d}{}^{f} F^{de} F_{e}{}^{h} \nabla_{c}F_{fh} + 
\frac{1}{2} dS_{a} dS^{a} dS^{b} F^{cd} \nabla_{d}F_{bc}\nn\\&& + 
\frac{1}{2} dS_{a} dS^{a} dS^{b} dS^{c} dS^{d} F_{b}{}^{e} 
\nabla_{d}F_{ce} + dS^{a} F_{b}{}^{d} F^{bc} F_{c}{}^{e} 
\nabla_{e}F_{ad} -  dS^{a} F_{a}{}^{b} F_{c}{}^{e} F^{cd} 
\nabla_{e}F_{bd} \nn\\&&+\! \frac{9}{20} dS_{a} dS^{a} dS_{b} dS^{b} dS^{c} 
F^{de} \nabla_{e}F_{cd}\! -\!  \frac{1}{2} dS^{a} F_{a}{}^{b} 
F_{b}{}^{c} F^{de} \nabla_{e}F_{cd}\! +\! \frac{9}{10} dS^{a} dS^{b} 
dS^{c} F_{a}{}^{d} F_{b}{}^{e} F_{d}{}^{f} \nabla_{e}F_{cf}\nn\\&& -  
\frac{1}{20} dS^{a} F_{a}{}^{b} F_{b}{}^{c} F_{c}{}^{d} F_{d}{}^{e} 
F^{fh} \nabla_{e}F_{fh} + \frac{1}{5} dS_{a} dS^{a} dS^{b} 
F_{c}{}^{e} F^{cd} F_{d}{}^{f} \nabla_{f}F_{be} \nn\\&&-  \frac{2}{5} 
dS^{a} dS^{b} dS^{c} F_{a}{}^{d} F_{b}{}^{e} F_{d}{}^{f} 
\nabla_{f}F_{ce} + \frac{1}{5} dS_{a} dS^{a} dS^{b} F_{b}{}^{c} 
F_{d}{}^{f} F^{de} \nabla_{f}F_{ce}\nn\\&& + \frac{4}{5} dS^{a} dS^{b} 
dS^{c} F_{a}{}^{d} F_{bd} F^{ef} \nabla_{f}F_{ce} -  \frac{1}{10} 
dS_{a} dS^{a} dS^{b} F_{b}{}^{c} F_{c}{}^{d} F^{ef} \nabla_{f}F_{de} \nn\\&&
-  \frac{6}{5} dS^{a} F_{b}{}^{d} F^{bc} F_{c}{}^{e} F_{d}{}^{f} 
F_{e}{}^{h} \nabla_{h}F_{af} + \frac{7}{40} dS^{a} F_{b}{}^{d} 
F^{bc} F_{c}{}^{e} F_{de} F^{fh} \nabla_{h}F_{af} \nn\\&&+ \frac{3}{10} 
dS^{a} F_{a}{}^{b} F_{c}{}^{e} F^{cd} F_{d}{}^{f} F_{e}{}^{h} 
\nabla_{h}F_{bf} -  \frac{1}{5} dS^{a} F_{a}{}^{b} F_{b}{}^{c} 
F_{c}{}^{d} F_{e}{}^{h} F^{ef} \nabla_{h}F_{df}\,,
\eeqa
\beqa
\delta A^a&\!\!\!\!\!\!\!=\!\!\!\!\!\!\!&\frac{4}{3} dS^{b} F_{b}{}^{c} \tilde{\Omega}^{a}{}_{c} -  
\frac{7}{40} dS^{b} F_{b}{}^{c} F_{d}{}^{f} F^{de} F_{e}{}^{h} 
F_{fh} \tilde{\Omega}^{a}{}_{c} -  \frac{21}{10} dS_{b} dS^{b} dS^{c} 
F_{c}{}^{d} \tilde{\Omega}^{a}{}_{d}\nn\\&& + \frac{247}{140} dS_{b} dS^{b} 
dS_{c} dS^{c} dS^{d} F_{d}{}^{e} \tilde{\Omega}^{a}{}_{e} + 
\frac{31}{30} dS^{b} F_{b}{}^{c} F_{c}{}^{d} F_{d}{}^{e} \tilde{\Omega}^{a}{}_{e} + \frac{17}{210} dS^{b} dS^{c} dS^{d} F_{b}{}^{e} 
F_{ce} F_{d}{}^{f} \tilde{\Omega}^{a}{}_{f}\nn\\&& - \! \frac{122}{105} dS_{b} 
dS^{b} dS^{c} F_{c}{}^{d} F_{d}{}^{e} F_{e}{}^{f} \tilde{\Omega}^{a}{}_{f} 
\!+ \!\frac{17}{70} dS^{b} F_{b}{}^{c} F_{c}{}^{d} F_{d}{}^{e} 
F_{e}{}^{f} F_{f}{}^{h} \tilde{\Omega}^{a}{}_{h} \!+\! \frac{7}{40} dS^{b} 
F^{ac} F_{d}{}^{f} F^{de} F_{e}{}^{h} F_{fh} \tilde{\Omega}_{bc}\nn\\&& -  
\frac{1}{2} dS^{b} F^{ac} F_{c}{}^{d} F_{d}{}^{e} \tilde{\Omega}_{be} + 
\frac{1}{10} dS^{b} F^{ac} F_{c}{}^{d} F_{d}{}^{e} F_{e}{}^{f} 
F_{f}{}^{h} \tilde{\Omega}_{bh} -  \frac{1}{2} dS^{b} dS^{c} dS^{d} 
F^{a}{}_{b} \tilde{\Omega}_{cd} \nn\\&&+ dS_{b} dS^{b} dS^{c} F^{ad} \tilde{\Omega}_{cd} -  \frac{1}{2} dS^{a} dS^{b} dS^{c} F_{b}{}^{d} \tilde{\Omega}_{cd} + \frac{3}{10} dS^{b} dS^{c} dS^{d} F^{ae} F_{b}{}^{f} 
F_{ef} \tilde{\Omega}_{cd}\nn\\&& -  dS^{b} F^{ac} F_{b}{}^{d} F_{d}{}^{e} 
\tilde{\Omega}_{ce} -  \frac{1}{5} dS_{b} dS^{b} dS^{c} F^{ad} 
F_{d}{}^{e} F_{e}{}^{f} \tilde{\Omega}_{cf} -  \frac{2}{5} dS^{a} dS^{b} 
dS^{c} F_{b}{}^{d} F_{d}{}^{e} F_{e}{}^{f} \tilde{\Omega}_{cf}\nn\\&& -  
\frac{1}{5} dS^{b} F^{ac} F_{b}{}^{d} F_{d}{}^{e} F_{e}{}^{f} 
F_{f}{}^{h} \tilde{\Omega}_{ch} + \frac{4}{5} dS_{b} dS^{b} dS^{c} 
dS^{d} dS^{e} F^{a}{}_{c} \tilde{\Omega}_{de} -  \frac{17}{20} dS_{b} 
dS^{b} dS_{c} dS^{c} dS^{d} F^{ae} \tilde{\Omega}_{de} \nn\\&&+ \frac{4}{5} 
dS^{a} dS_{b} dS^{b} dS^{c} dS^{d} F_{c}{}^{e} \tilde{\Omega}_{de} + 
dS^{b} F^{ac} F_{b}{}^{d} F_{c}{}^{e} \tilde{\Omega}_{de} -  \frac{1}{2} 
dS^{b} dS^{c} dS^{d} F^{ae} F_{b}{}^{f} F_{cf} \tilde{\Omega}_{de}\nn\\&& -  
dS^{b} F^{a}{}_{b} F_{c}{}^{e} F^{cd} \tilde{\Omega}_{de} + \frac{3}{10} 
dS^{b} dS^{c} dS^{d} F^{ae} F_{be} F_{c}{}^{f} \tilde{\Omega}_{df} -  
\frac{9}{10} dS^{b} dS^{c} dS^{d} F^{a}{}_{b} F_{c}{}^{e} 
F_{e}{}^{f} \tilde{\Omega}_{df} \nn\\&&+ \frac{6}{5} dS_{b} dS^{b} dS^{c} 
F^{ad} F_{c}{}^{e} F_{e}{}^{f} \tilde{\Omega}_{df} + \frac{1}{5} dS^{b} 
F^{ac} F_{b}{}^{d} F_{c}{}^{e} F_{e}{}^{f} F_{f}{}^{h} \tilde{\Omega}_{dh} 
+ \frac{1}{10} dS^{b} dS^{c} dS^{d} F^{a}{}_{b} F_{c}{}^{e} 
F_{d}{}^{f} \tilde{\Omega}_{ef}\nn\\&& -  \frac{3}{10} dS_{b} dS^{b} dS^{c} 
F^{ad} F_{c}{}^{e} F_{d}{}^{f} \tilde{\Omega}_{ef} -  \frac{2}{5} dS^{a} 
dS^{b} dS^{c} F_{b}{}^{d} F_{c}{}^{e} F_{d}{}^{f} \tilde{\Omega}_{ef} + 
\frac{1}{5} dS_{b} dS^{b} dS^{c} F^{a}{}_{c} F_{d}{}^{f} F^{de} 
\tilde{\Omega}_{ef}\nn\\&& - \! \frac{7}{10} dS^{b} F^{ac} F_{b}{}^{d} 
F_{c}{}^{e} F_{d}{}^{f} F_{f}{}^{h} \tilde{\Omega}_{eh}\! +\! \frac{7}{10} 
dS^{b} F^{ac} F_{b}{}^{d} F_{c}{}^{e} F_{d}{}^{f} F_{e}{}^{h} \tilde{\Omega} _{fh} \!- \! \frac{1}{5} dS^{b} F^{a}{}_{b} F_{c}{}^{e} F^{cd} 
F_{d}{}^{f} F_{e}{}^{h} \tilde{\Omega}_{fh}\nn\\&& -  \frac{4}{5} dS^{b} F^{ac} 
F_{b}{}^{d} F_{cd} F_{e}{}^{h} F^{ef} \tilde{\Omega}_{fh} -  
\frac{7}{40} dS^{b} dS^{c} F_{d}{}^{f} F^{de} F_{e}{}^{h} F_{fh} 
\nabla_{c}F^{a}{}_{b} + \frac{7}{30} dS^{b} dS^{c} F_{b}{}^{d} 
F_{d}{}^{e} \nabla_{c}F^{a}{}_{e} \nn\\&&-  \frac{23}{70} dS^{b} dS^{c} 
F_{b}{}^{d} F_{d}{}^{e} F_{e}{}^{f} F_{f}{}^{h} 
\nabla_{c}F^{a}{}_{h} -  dS^{b} dS^{c} F^{ad} F_{d}{}^{e} 
\nabla_{c}F_{be} -  \frac{1}{5} dS^{b} dS^{c} F^{ad} F_{d}{}^{e} 
F_{e}{}^{f} F_{f}{}^{h} \nabla_{c}F_{bh}\nn\\&& -  \frac{1}{2} dS_{b} 
dS^{b} dS^{c} dS^{d} \nabla_{d}F^{a}{}_{c} -  \frac{2}{105} dS_{b} 
dS^{b} dS^{c} dS^{d} F_{c}{}^{e} F_{e}{}^{f} \nabla_{d}F^{a}{}_{f} + 
dS^{b} dS^{c} F^{ad} F_{b}{}^{e} \nabla_{d}F_{ce}\nn\\&& + \frac{3}{10} 
dS_{b} dS^{b} dS^{c} dS^{d} F^{ae} F_{e}{}^{f} \nabla_{d}F_{cf} -  
\frac{1}{10} dS^{a} dS^{b} dS^{c} dS^{d} F_{b}{}^{e} F_{e}{}^{f} 
\nabla_{d}F_{cf} \nn\\&&+ \frac{1}{5} dS^{b} dS^{c} F^{ad} F_{b}{}^{e} 
F_{e}{}^{f} F_{f}{}^{h} \nabla_{d}F_{ch} + \frac{23}{30} dS^{b} 
dS^{c} F_{b}{}^{d} F_{d}{}^{e} \nabla_{e}F^{a}{}_{c} \nn\\&& +\! 
\frac{1}{20} dS_{b} dS^{b} dS_{c} dS^{c} dS^{d} dS^{e} 
\nabla_{e}F^{a}{}_{d}\!-\!  \frac{1}{2} dS^{b} dS^{c} F_{b}{}^{d} 
F_{c}{}^{e} \nabla_{e}F^{a}{}_{d}\! \!- \! \frac{3}{10} dS^{b} dS^{c} 
dS^{d} dS^{e} F_{b}{}^{f} F_{cf} \nabla_{e}F^{a}{}_{d} \nn\\&&+ 
\frac{3}{14} dS^{b} dS^{c} F_{b}{}^{d} F_{c}{}^{e} F_{d}{}^{f} 
F_{f}{}^{h} \nabla_{e}F^{a}{}_{h} -  dS^{a} dS^{b} F_{c}{}^{e} 
F^{cd} \nabla_{e}F_{bd} -  \frac{1}{2} dS^{b} dS^{c} F^{a}{}_{b} 
F^{de} \nabla_{e}F_{cd} \nn\\&&-  \frac{1}{2} dS^{a} dS^{b} F_{b}{}^{c} 
F^{de} \nabla_{e}F_{cd} + \frac{2}{5} dS^{b} dS^{c} dS^{d} dS^{e} 
F^{a}{}_{b} F_{c}{}^{f} \nabla_{e}F_{df} -  \frac{6}{5} dS_{b} 
dS^{b} dS^{c} dS^{d} F^{ae} F_{c}{}^{f} \nabla_{e}F_{df}\nn\\&& -  
\frac{82}{105} dS_{b} dS^{b} dS^{c} dS^{d} F_{c}{}^{e} F_{e}{}^{f} 
\nabla_{f}F^{a}{}_{d} + \frac{2}{5} dS_{b} dS^{b} dS^{c} dS^{d} 
F_{c}{}^{e} F_{d}{}^{f} \nabla_{f}F^{a}{}_{e}\nn\\&& -  \frac{7}{20} 
dS_{b} dS^{b} dS_{c} dS^{c} F_{d}{}^{f} F^{de} \nabla_{f}F^{a}{}_{e} 
+ \frac{1}{5} dS^{a} dS_{b} dS^{b} dS^{c} F_{d}{}^{f} F^{de} 
\nabla_{f}F_{ce}\nn\\&& -  \frac{7}{10} dS^{b} dS^{c} F^{ad} F_{b}{}^{e} 
F_{d}{}^{f} F_{e}{}^{h} \nabla_{f}F_{ch} + \frac{1}{2} dS^{a} 
dS^{b} dS^{c} dS^{d} F_{b}{}^{e} F_{c}{}^{f} \nabla_{f}F_{de}\nn\\&& -  
\frac{1}{10} dS_{b} dS^{b} dS^{c} dS^{d} F^{a}{}_{c} F^{ef} 
\nabla_{f}F_{de} -  \frac{7}{20} dS_{b} dS^{b} dS_{c} dS^{c} 
F^{ad} F^{ef} \nabla_{f}F_{de}\nn\\&& -  \frac{1}{10} dS^{a} dS_{b} 
dS^{b} dS^{c} F_{c}{}^{d} F^{ef} \nabla_{f}F_{de} -  \frac{1}{5} 
dS^{b} dS^{c} F^{ad} F_{b}{}^{e} F_{c}{}^{f} F_{e}{}^{h} 
\nabla_{f}F_{dh}\nn\\&& -  \frac{7}{10} dS^{b} dS^{c} F^{ad} F_{b}{}^{e} 
F_{c}{}^{f} F_{d}{}^{h} \nabla_{f}F_{eh} + \frac{9}{70} dS^{b} 
dS^{c} F_{b}{}^{d} F_{d}{}^{e} F_{e}{}^{f} F_{f}{}^{h} 
\nabla_{h}F^{a}{}_{c}\nn\\&& -  \frac{1}{70} dS^{b} dS^{c} F_{b}{}^{d} 
F_{c}{}^{e} F_{d}{}^{f} F_{f}{}^{h} \nabla_{h}F^{a}{}_{e} -  
\frac{1}{10} dS^{b} dS^{c} F_{b}{}^{d} F_{c}{}^{e} F_{d}{}^{f} 
F_{e}{}^{h} \nabla_{h}F^{a}{}_{f}\nn\\&& -  \frac{7}{10} dS^{b} dS^{c} 
F_{b}{}^{d} F_{cd} F_{e}{}^{h} F^{ef} \nabla_{h}F^{a}{}_{f} + 
\frac{3}{10} dS^{a} dS^{b} F_{c}{}^{e} F^{cd} F_{d}{}^{f} 
F_{e}{}^{h} \nabla_{h}F_{bf} \nn\\&&-  \frac{7}{10} dS^{b} dS^{c} F^{ad} 
F_{b}{}^{e} F_{d}{}^{f} F_{f}{}^{h} \nabla_{h}F_{ce} + \frac{7}{5} 
dS^{b} dS^{c} F^{ad} F_{b}{}^{e} F_{d}{}^{f} F_{e}{}^{h} 
\nabla_{h}F_{cf}\nn\\&& -  \frac{6}{5} dS^{b} dS^{c} F^{a}{}_{b} 
F_{d}{}^{f} F^{de} F_{e}{}^{h} \nabla_{h}F_{cf} -  \frac{1}{5} 
dS^{a} dS^{b} F_{b}{}^{c} F_{d}{}^{f} F^{de} F_{e}{}^{h} 
\nabla_{h}F_{cf} \nn\\&&-  \frac{4}{5} dS^{b} dS^{c} F^{ad} F_{bd} 
F_{e}{}^{h} F^{ef} \nabla_{h}F_{cf} -  \frac{3}{5} dS^{b} dS^{c} 
F^{ad} F_{b}{}^{e} F_{de} F^{fh} \nabla_{h}F_{cf} \nn\\&&+ \frac{1}{5} 
dS^{b} dS^{c} F^{ad} F_{b}{}^{e} F_{c}{}^{f} F_{e}{}^{h} 
\nabla_{h}F_{df} + \frac{4}{5} dS^{b} dS^{c} F^{a}{}_{b} 
F_{c}{}^{d} F_{e}{}^{h} F^{ef} \nabla_{h}F_{df}\nn\\&& -  \frac{1}{5} 
dS^{a} dS^{b} F_{b}{}^{c} F_{c}{}^{d} F_{e}{}^{h} F^{ef} 
\nabla_{h}F_{df} -  \frac{7}{10} dS^{b} dS^{c} F^{ad} F_{b}{}^{e} 
F_{ce} F^{fh} \nabla_{h}F_{df}\nn\\&& -  \frac{3}{5} dS^{b} dS^{c} F^{ad} 
F_{bd} F_{c}{}^{e} F^{fh} \nabla_{h}F_{ef} + \frac{3}{5} dS^{b} 
dS^{c} F^{a}{}_{b} F_{c}{}^{d} F_{d}{}^{e} F^{fh} \nabla_{h}F_{ef}\nn\\&& + 
\frac{1}{10} dS^{a} dS^{b} F_{b}{}^{c} F_{c}{}^{d} F_{d}{}^{e} 
F^{fh} \nabla_{h}F_{ef}\,,
\eeqa
where all world-volume indices are the indices in the base space, e.g. $a=\ta$, $b=\ta$, and so on. In the above equations, $\tilde{\Omega}_{\ta\tb} = \tilde{\Omega}_{\ta\tb}{}^y$.
%\newpage

\end{document}